\documentclass[a4paper,12pt]{article}
\usepackage{amsmath,amssymb,amsthm}

\newcommand{\<}{\langle}
\newcommand{\id}{\operatorname{Id}}
\renewcommand{\>}{\rangle}
 
\newtheorem{remark}{Remark}
 
\title{Operator perturbation weighing, from quasi-critical source
system response}

\author{Pierre Albar\`ede\\
\small r\'es.  Valvert, 12 rue de la Fourane, 13090 
Aix-en-Provence, France
\\\small palbarede@yahoo.com}

\date{\today}
\begin{document}
\maketitle

\begin{abstract}
    In Hilbert space, a linear source-to-flux problem in the critical
    (zero eigenvalue) limit is ill-posed, but regularized by a
    constraint on a linear functional, fulfilled by tuning some
    control variable.  For any exciting perturbation, I obtain, by
    spectral decomposition and perturbation theory, the regularized
    flux and the regularizing control variable non-linear responses.
    
    May the exciting perturbation be obtained, inversely, from
    observable responses?  Yes, in some cases, from the existence of
    a weight scale, a perturbation series, determined by recursion
    relations, involving well-posed source problems, and the
    possibility of obtaining this weight scale from observables of
    both the unconstrained and constrained systems.  \vspace{1em}

\noindent key words: source, non-linear response, critical,
regularization, perturbation, self-shielding, operator weighing,
observable.  \vspace{1em}


\end{abstract}
% MSC2000
% 41 APPROXIMATIONS AND EXPANSIONS
% 41A29 Approximation with constraints
% 47 OPERATOR THEORY
% 47A52 Ill-posed problems, regularization
% 47A55 Perturbation theory
% 47A60 Functional calculus
% 82C70 Transport processes
% 93 SYSTEMS THEORY; CONTROL
% 93B05 Controllability 
% 93B07 Observability 
% 93B28 Operator-theoretic methods 
% 93B30 System identification 
% 93C05 Control systems, guided systems/linear systems
% 93D15 Stabilization of systems by feedback
% 
% PACS
% 02.10.Sp Linear and multilinear algebra; matrix theory (finite and infinite)
% 02.30.Wd Calculus of variations and optimal control
% 02.30.Px Approximations and expansions
% 02.30.Tb Operator theory
% 05.60.+w Transport processes: theory
% 06.20.Dk Measurement and error theory
% 07.10.Lw Balance systems, tensile machines, etc.

\newpage
%LANL math-ph/0007026 Password: kn3gg
\tableofcontents
\newpage
\section{Problem setup\label{setup}}

In $\mathcal{H}$, a real Hilbert space, with the scalar product
$\<.,.\>$ and the norm
\mbox{$\parallel\psi\parallel\equiv\<\psi,\psi\>^{1/2}$},
let $L:\mathcal{F}\rightarrow\mathcal{Q}\equiv L(\mathcal{F})$ be a linear
(additive and homogeneous) endomorphism
($\mathcal{F},\mathcal{Q}\subset\mathcal{H}$ are linear manifolds).

I consider the source problem, finding, for some source
$Q\in\mathcal{Q}$, the flux $\psi\in\mathcal{F}$ such that
\begin{equation}
    0=L\psi+Q.
    \label{0=LP+Q}
\end{equation}
I assume that the `flux-to-source' operator $L$ is invertible (on the
left), hence
\begin{equation}
    \psi=\Phi(L,Q)\equiv-L^{-1}Q.
    \label{?P}
\end{equation}

It occurs that the flux operator $\Phi$, defined in (\ref{?P}), is
linear in $Q$, but \emph{non-linear} in $L$.  This non-linearity is
formally eliminated by the change of variable $L\mapsto L^{-1}$. 
However, expressing an inverse operator is often difficult, in
particular, when the operator $L$ has an eigenvalue close to zero (in
this case, the non-linearity is somehow traded for a singularity). 
Therefore, I will keep the variable $L$, and the non-linearity
$L\mapsto\Phi$.

The small eigenvalue \emph{critical limit} of the source problem is
conveniently defined by the following hypotheses:
\begin{itemize}
    \item $L,Q$ are functions of some scalar variable $z$
    (varying over an interval);

    \item $\sigma(z)$ is a single eigenvalue of $L(z)$ associated with
    an eigenvector $\varphi(z)$;

    \item $z_{c}$ is a critical value, such that
\begin{equation}
    \lim_{z\rightarrow z_{c}}\sigma(z)=0;
    \label{?crilim}
\end{equation}

    \item for all $z\neq z_{c}$, $L(z)$ has an inverse on the left
    $L^{-1}(z)$.
\end{itemize}

Even though the source problem has no more than one solution for
$z\neq z_{c}$, the source problem (\ref{0=LP+Q}) is still ill-posed in
the critical limit $z\rightarrow z_{c}$, because $\Phi$ is infinitely
sensitive to perturbations of $Q$ (essentially:
$z\mapsto\sigma(z)^{-1}$ has no derivative at $z_{c}$).

I take for regularizing condition the permanence of some linear
continuous `gauge' functional $Q^{\dag}$ of flux.  This will be
achieved by tuning the variable $z$, thus appearing as a \emph{control
variable}.  Equivalently, in the language of system analysis, a
high-gain amplifier can be hardly used without feedback control, in
order to maintain a reasonable output level (not too large, not too
small).  As the Hilbert space is self-dual \cite[\S 9.5, p. 
264]{Vulikh}, the regularizing condition expresses as the constraint
\begin{equation}
    R\equiv\<Q^{\dag},\Phi\>=R_{0},Q^{\dag}\in\mathcal{H}.
    \label{constraint}
\end{equation}

The system parameters are $T\equiv(L,Q,Q^{\dag})$, all functions of
$z$.  $T$ may be excited by some perturbation $\delta T$, actually
driven by some exciting variable $\epsilon$, independent of $z$.

Primary \emph{observables} are the scalars outputs of the system, here
\begin{itemize}
    \item  the exciting variable $\epsilon$,

    \item  the control variable $z$,

    \item the gauge output $R$.
\end{itemize}
Secondary observables are parametric expressions thereof, independent
of the system parameters $T,\delta T$, for example, $\ln R$, or the
partial derivative $(\partial R/\partial z)_{\epsilon}$.

Observation is useless, without an \emph{interpretation relation},
between $T,\delta T$ and observables.  The main issue of the paper is
to prove the existence of an interpretation relation, of the form
\begin{equation}
    \boxed{Z_{1}(T,\delta T)+Z_{2}(\text{prim. observables})=0},
    \label{Z1+Z2=0}
\end{equation}
where $Z_{1},Z_{2}$ are scalar \emph{weight} functionals.  The weight
$Z_{1}$ is obtained by processing the primary observables through
$Z_{2}$ and (\ref{Z1+Z2=0}).  I will explicitly construct
$Z_{1},Z_{2}$.

Ideally, the sources are unexcited ($0=\delta Q=\delta Q^{\dag}$) and
the exciting perturbation is uncontrolled ($\delta L$ does not depend
on $z$).  Solving (\ref{Z1+Z2=0}) for $\delta L$, if possible, is the
principle of a direct operator perturbation measurement method.

\section{The source system and the critical limit}

\subsection{Unconstrained system definition and
properties\label{sysdef}}

The source system is represented by (\ref{0=LP+Q}).  As the constraint
is not used, the control variable $z$ is left constant and implicit
(`asleep') until the critical limit is considered (section
\ref{crilim}).  In the language of system analysis, the control loop
is open.

The gauge output is
\begin{equation}
    R(T)\equiv\<Q^{\dag},\Phi(L,Q)\>,Q^{\dag}\in\mathcal{H}.
    \label{?R}
\end{equation}

I assume that $L$ has an adjoint
$L^{\dag}:\mathcal{F}^{\dag}\rightarrow\mathcal{Q}^{\dag}\equiv
L(\mathcal{F}^{\dag})$ and $Q^{\dag}\in\mathcal{Q}^{\dag}$.  The
adjoint source system is
\begin{equation}
    0=L^{\dag}\psi^{\dag}+Q^{\dag}.
    \label{(0=LP+Q)^+}
\end{equation}

I assume that $L^{\dag}$ has an inverse 
$L^{\dag-1}:\mathcal{Q}^{\dag}\rightarrow\mathcal{F}^{\dag}$, solving 
the source problem (\ref{(0=LP+Q)^+}):
\begin{equation}
    \forall Q^{\dag}\in\mathcal{Q}^{\dag},
    \psi^{\dag}=\Phi^{\dag}(L,Q^{\dag})\equiv-L^{\dag-1}Q^{\dag}=
    \Phi(L^{\dag},Q^{\dag}).
    \label{?P^+}
\end{equation}

I introduce a restriction of the scalar product, noted $\<.|.\>$, to
\[(\mathcal{Q^{\dag}}\times\mathcal{F})
\cup(\mathcal{F}^{\dag}\times\mathcal{Q}).\] Considering, adjointly,
the source as the gauge functional of the adjoint source system, the
adjoint gauge output is
\begin{equation}
    R^{\dag}(T^{\dag})\equiv\<\Phi^{\dag}(L,Q^{\dag})|Q\>,\ 
    T^{\dag}\equiv(L,Q^{\dag},Q)
    \label{?R^+}
\end{equation}
One shows that $L^{\dag-1}$ is adjoint to $L^{-1}$:
\begin{equation}
    L^{\dag-1}=L^{-1\dag}.
    \label{/L^+}
\end{equation}
Using (\ref{/L^+}) in (\ref{?R}) yields the reciprocity relation 
\cite{Bell},
\begin{equation}
    R(T)=\<Q^{\dag}|\Phi(L,Q)\>=\<\Phi^{\dag}(L,Q^{\dag})|Q\>=
    R^{\dag}(T^{\dag}).
    \label{R=R^+}
\end{equation}
From (\ref{?P^+}, \ref{R=R^+}), the adjoint flux
$\Phi^{\dag}(L,Q^{\dag})$, considered as a functional of $Q$, produces
the gauge output.

The linearity of the flux-to-source operator and the gauge functional 
imply \emph{scaling laws}:
\begin{equation}
    \forall\alpha\in\mathbb{R}^{*},
    \alpha R_{0}=\<\alpha Q^{\dag}|\Phi(L,Q)\>=
    \<\alpha\Phi^{\dag}(L,Q^{\dag})|Q\>.
    \label{gaugeinv}
\end{equation}
Considering the source and the flux as the physical quantities, while 
the adjoint flux and the gain factor $\alpha$ are mathematical and 
operational commodities, (\ref{gaugeinv}) shows that the physical 
quantities are independent of $\alpha$, which corresponds to what 
physicists call \emph{gauge invariance} \cite[\S VII.5]{Brillouin}.

The unconstrained system is represented by the parametric expression
\begin{equation}
    U(T)\equiv(R,\Phi,\Phi^{\dag})(T), 
    \label{?U}
\end{equation}
mapping the independent variables, or parameters, onto the dependent 
variables, the gauge output and the fluxes.  With (\ref{?U}), 
(\ref{gaugeinv}) becomes
\begin{equation}
    \forall\alpha\in\mathbb{R}^{*},U(L,Q,\alpha Q^{\dag})
    =(\alpha R,\Phi,\alpha\Phi^{\dag}).
    \label{scaling1}
\end{equation}

By definition (\ref{?P}), and adjointly, the direct and adjoint source 
systems are uncoupled:
\begin{equation}
    0=\frac{\partial\Phi}{\partial Q^{\dag}}
    =\frac{\partial\Phi^{\dag}}{\partial Q}.
    \label{dP/dQ^+}
\end{equation}

\subsection{The spectral decomposition\label{funhardec}}

I assume that $L^{\dag}$ has an eigenvalue $\sigma^{\dag}$, and that
the associated eigenspace is of dimension one and eigenvector
$\varphi^{\dag},\<\varphi^{\dag},\varphi\>\neq 0$.  Hence, the
eigenvectors can be and are normalized, so that
\begin{equation}
    \<\varphi^{\dag}|\varphi\>=1.
    \label{<p,p>=1}
\end{equation}
(As $\sigma\neq 0$, $\varphi\in\mathcal{Q}$, and the use of the 
restricted product above is correct.)  One shows (classically)
\begin{equation}
    0=(\sigma^{\dag}-\sigma)\<\varphi^{\dag}|\varphi\>,
    \nonumber
\end{equation}
hence, from (\ref{<p,p>=1}), $\sigma^{\dag}=\sigma$.  One shows
that $\tilde{\mathcal{F}}\equiv\varphi^{\dag\perp}\cap\mathcal{F}$ is
stable for $L$, and, adjointly,
$\widetilde{\mathcal{F}^{\dag}}\equiv\varphi^{\perp}\cap\mathcal{F}^{\dag}$
is stable for $L^{\dag}$.

$\varphi^{\dag\perp}$ is the hyper-plane, orthogonal to
$\varphi^{\dag}$, a subspace of codimension one.  From
(\ref{<p,p>=1}), $\varphi\notin\varphi^{\dag\perp}$, so that
$\mathcal{H}=\varphi\oplus\varphi^{\dag\perp}$ and one shows
\begin{equation}
    \mathcal{F}=\varphi\oplus\tilde{\mathcal{F}},
    \mathcal{F}^{\dag}=\varphi^{\dag}\oplus\widetilde{\mathcal{F}^{\dag}},
    \nonumber
\end{equation}
which are equivalent to `closure relations',
\begin{equation}
    1_{\mathcal{F}}=\varphi\varphi^{\dag}+\tilde{\pi}, 
    1_{\mathcal{F}^{\dag}}=\varphi^{\dag}\varphi+\tilde{\pi}^{\dag}, 
    \label{closurep}
\end{equation}
where 
\begin{itemize}
    \item $\varphi\varphi^{\dag}$ stands for the tensor product 
    $|\varphi\>\negmedspace\<\varphi^{\dag}|$,

    \item $\varphi^{\dag}\varphi$ stands for the tensor product 
    $|\varphi^{\dag}\>\negmedspace\<\varphi|$, adjoint to the latter,

    \item $\tilde{\pi}$ is the projection, in $\mathcal{F}$, on 
    $\tilde{\mathcal{F}}$, along $\varphi$,

    \item $\tilde{\pi}^{\dag}$ is the projection, in 
    $\mathcal{F}^{\dag}$, on $\widetilde{\mathcal{F}^{\dag}}$, along 
    $\varphi^{\dag}$, adjoint to the latter.
\end{itemize}
Condensed notations will be necessary:
\begin{equation}
    \forall\psi\in\mathcal{F},\tilde{\psi}\equiv\tilde{\pi}\psi, 
    \forall\psi^{\dag}\in\mathcal{F}^{\dag}, 
    \widetilde{\psi^{\dag}}\equiv\tilde{\pi}^{\dag}\psi^{\dag}, 
    \tilde{L}\equiv\tilde{\pi}L, 
    \widetilde{L^{\dag}}\equiv\tilde{\pi}^{\dag}L^{\dag}.  
    \nonumber
\end{equation}

Applying $\tilde{\pi}$ to (\ref{0=LP+Q}), 
\begin{equation}
    0=\tilde{L}\psi+\tilde{Q}=\tilde{L}\tilde{\psi}+\tilde{Q}, 
    \label{0=TiLTiP+TiQ}
\end{equation} 
which is solved for the harmonic flux:
\begin{equation}
    \tilde{\Phi}=-\tilde{L}^{-1}\tilde{Q}, 
    \label{TiP=-TiQ/TiL}
\end{equation}
where $\tilde{L}$ has been restricted to the stable subspace 
$\tilde{\mathcal{F}}$.  Applying $\<\varphi^{\dag}|$ to 
(\ref{0=LP+Q}),
\begin{equation}
    \<\varphi^{\dag}|\Phi\>=-\frac{\<\varphi^{\dag}|Q\>}{\sigma}.
    \label{<p,P>=-<p,Q>/s}
\end{equation}

Combining (\ref{closurep}, \ref{<p,P>=-<p,Q>/s}, \ref{TiP=-TiQ/TiL}), 
\begin{eqnarray}
    \Phi&=&-\frac{\<\varphi^{\dag}|Q\>}{\sigma}\varphi
    -\tilde{L}^{-1}\tilde{Q}, 
    \label{P=-<p,Q>p/s}\\
    R&=&-\frac{\<Q^{\dag}|\varphi\varphi^{\dag}|Q\>}{\sigma}
    +\<Q^{\dag}|-\tilde{L}^{-1}\tilde{Q}\>.
    \label{R=-<Q,pp,Q>/s}
\end{eqnarray}
On (\ref{P=-<p,Q>p/s}) appears the (unconstrained) source problem
critical ill-posedness.  Eliminating $\sigma$ from
(\ref{R=-<Q,pp,Q>/s}) and inserting the result into
(\ref{P=-<p,Q>p/s}),
\begin{eqnarray}
    \omega&\equiv&\frac{\<Q^{\dag}|-\tilde{L}^{-1}\tilde{Q}\>}{R}
    =\frac{\<Q^{\dag}|\tilde{\Phi}\>}{\<Q^{\dag}|\Phi\>}, 
    \label{?Tiom}\\    
    \sigma&=&-\frac{\<Q^{\dag}|\varphi\varphi^{\dag}|Q\>}
    {R(1-\omega)}, 
    \label{s=}\\
    \Phi&=&\frac{R(1-\omega)}
    {\<Q^{\dag}|\varphi\>}\varphi-\tilde{L}^{-1}\tilde{Q}.
    \label{P=R/<Q,p>}
\end{eqnarray}
$\omega$ is the `harmonicity'.  The negative source-to-flux operator 
is
\begin{equation}
    -L^{-1}=\frac{R(1-\omega)}{\<Q^{\dag}|\varphi\varphi^{\dag}|Q\>}
    \varphi\varphi^{\dag}-\tilde{L}^{-1}\tilde{\pi}.
    \label{-/L=}
\end{equation}

Some useful relations are, adjointly,
\begin{eqnarray}
    \Phi^{\dag}&=&
    \frac{R(1-\omega^{\dag})}{\<\varphi^{\dag}|Q\>}\varphi^{\dag}
    -\widetilde{L^{\dag}}^{-1}\tilde{Q^{\dag}}, 
    \label{P^=R/<p,Q>}\\
    \omega^{\dag}&\equiv&
    \frac{\<-\widetilde{L^{\dag}}^{-1}\widetilde{Q^{\dag}}|Q\>}{R}
    =\frac{\<\widetilde{\Phi^{\dag}}|Q\>}{\<\Phi^{\dag}|Q\>}=\omega,
    \nonumber
\end{eqnarray} 
and
\begin{eqnarray}    
    \tilde{\pi}^{2}&=&\tilde{\pi},
    \nonumber\\
    \tilde{\pi}L&=&L\tilde{\pi},
    \nonumber\\
    \widetilde{L^{-1}}&=&\tilde{L}^{-1},
    \nonumber\\
    \<Q^{\dag}|-\tilde{L}^{-1}\tilde{\pi}&=&\<\Phi^{\dag}|\tilde{\pi},
    \label{algTi}
\end{eqnarray}
and adjointly.

\subsection{The constant gauge output critical limit\label{crilim}}

I show up the (smooth) scalar control variable $z$.  Acknowledging the
functional character of the parameters $T$, (\ref{?U}) must be
corrected to show chained functions:
\begin{equation}
    U\circ T=(R,\Phi,\Phi^{\dag})\circ T.
    \nonumber
\end{equation}

I assume that observables are bounded in the critical limit
(\ref{?crilim}).  Considering moreover (\ref{s=}, \ref{P=R/<Q,p>},
\ref{P^=R/<p,Q>}), the direct flux or the adjoint flux must diverge.

I assume
\begin{eqnarray}
    \<\varphi^{\dag}|Q\>\varphi(z)=\mathcal{O}(\sigma(z)),
    z\rightarrow z_{c},
    \nonumber\\
    \exists m_{1},\forall z,
    0<m_{1}\le\parallel\<Q^{\dag}|\varphi\>\varphi^{\dag}(z)\parallel,
    \label{Q->0}
\end{eqnarray}
(while the source is turned off, the gauge is not) so that the direct
flux does not diverge and the adjoint flux does, though only by a
scalar factor.  Following the comments of section \ref{sysdef} on
gauge invariance, the direct flux boundedness is physically motivated.

I assume spectral separation, uniform in $z$:
\begin{multline}
    \exists m,\forall z,
    \forall(\tilde{\psi}\in\tilde{\mathcal{F}},
    \parallel\tilde{\psi}\parallel=1),
    \forall(\widetilde{\psi^{\dag}}\in\widetilde{\mathcal{F}^{\dag}},
    \parallel\widetilde{\psi^{\dag}}\parallel=1),\\
    0<m\le\min(\parallel\tilde{L}(z)\tilde{\psi}\parallel, 
    \parallel\widetilde{L^{\dag}}(z)\widetilde{\psi^{\dag}}\parallel).
    \label{spsep}
\end{multline}
Consequently, any eigenvalue of $\tilde{L}$ has a modulus greater than
$m$.  Only the eigenvalue $\sigma(z)$ gets close to zero.  $\sigma(z)$
and $\varphi(z)$ are the `fundamental' eigenvalue and eigenvector
(and adjointly).  The maximum of all $m$ in (\ref{spsep}) is the
`spectral gap'.  The harmonicity and the ratios of the second terms
over the first terms in the r.  h. s.  of (\ref{P=R/<Q,p>},
\ref{P^=R/<p,Q>}) are $\mathcal{O}(\sigma(z)/m)$.  The constant gauge
output critical limits are
\begin{eqnarray}
    \omega\circ T(z)&=&\mathcal{O}(\sigma(z)/m), 
    \label{Tiom->0}\\
    \Phi\circ T(z)&\sim&\frac{R_{0}}
    {\<Q^{\dag}|\varphi_{c}\>(z_{c})}\varphi_{c},\ 
    \varphi_{c}\equiv\varphi(z_{c}),
    \nonumber\\
    \Phi^{\dag}\circ T(z)&\sim&\frac{R_{0}}
    {\<\varphi^{\dag}_{c}|Q\>(z_{c})}\varphi^{\dag}_{c},\  
    \varphi^{\dag}_{c}\equiv\varphi^{\dag}(z_{c}).
    \nonumber
\end{eqnarray}

\section{Perturbation theory with constraint\label{PT}}

\subsection{The constraint operator}

The constraint consists in tuning the control variable $z$ to the 
`balancing' value $\underline{z}$, such that the gauge output remains 
equal to a reference value $R_{0}$, which is a new independent 
variable, while $\underline{z}$ is a new dependent variable. 

I assume that the problem has a unique solution:
\begin{equation}
    \exists!\underline{z}(T,R_{0}),R(T(\underline{z}))=R_{0}.    
    \label{?z_}
\end{equation}
\begin{remark}
    As $T$ is a function (of $z$), $\underline{z}(.,R_{0})$ is a 
    (non-linear) functional (of the function $T$).
\end{remark}

For all function $t$ of $z$, like $T$, is defined the constrained 
value
\begin{equation}
    \underline{t}(T,R_{0})\equiv t(\underline{z}(T,R_{0})).
    \nonumber
\end{equation}
The (linear) `constraint operator', 
\begin{equation}
    \_:t\mapsto\underline{t}=t\circ\underline{z}, 
    \label{?_1}
\end{equation}
transforms a function of $z$ into an operator on $(T, R_{0})$.  
(\ref{?_1}) is formally the commutation relation
\begin{equation}
    \_t=t\_. 
    \label{_f=f_}
\end{equation} 
For all operator $u$ on $T(z)$ (like $\id,U$),
$\underline{u}(T)\equiv\underline{u\circ T}=u(\underline{T})
\triangleright\underline{u}$.  The symbol $\triangleright$ introduces
abbreviations (to be used with care).

The constrained system is thus represented by $(\underline{U}, 
\underline{z})$, where $\underline{\Phi}$ should be evaluated from 
(\ref{P=R/<Q,p>}), where $R\rightarrow R_{0}$.  On the constrained 
system, (\ref{scaling1}) becomes
\begin{equation}
    \forall\alpha\in\mathbb{R}^{*},
    (\underline{U},\underline{z})
    ((L,Q,\alpha Q^{\dag}),\alpha R_{0})
    =((\alpha\underline{R},\underline{\Phi},\alpha\underline{\Phi}^{\dag}), 
    \underline{z})((L,Q,Q^{\dag}),R_{0})). 
    \label{scaling2}
\end{equation}

The adjoint source $Q^{\dag}$, although it does not appear in 
(\ref{0=LP+Q}), nevertheless affects the flux, because the 
flux-to-source operator depends on $\underline{z}$, that depends on 
$Q^{\dag}$; the constraint couples the direct and adjoint source 
systems; as opposed to (\ref{dP/dQ^+}),
\begin{equation}
    \frac{\partial\underline{\Phi}}{\partial Q^{\dag}}\neq 0,
    \frac{\partial\underline{\Phi}^{\dag}}{\partial Q}\neq 0.
    \label{dP/dQ^+_}
\end{equation}

\subsection{Functional variational definitions\label{pervardef}}

The perturbation of any vector $x$ is noted, as usual,
\begin{equation*}
    (x,\delta x)^{*}\equiv x+\delta x\triangleright x^{*}.
%    \label{?x*}
\end{equation*}
$x,x^{*}$ are respectively the reference and perturbed values.  The
definition is extended to any parametric expression $f(x)$:
\begin{eqnarray}
    (f(x), \delta f, \delta x)^{*}&\equiv&(f+\delta f)(x+\delta x)
    \triangleright f^{*}(x^{*})\triangleright f(x)^{*}, 
    \label{?f(x)*}\\
    \delta(f(x), \delta f, \delta x)&\equiv&
    (f+\delta f)(x+\delta x)-f(x)\triangleright\delta(f(x))
    \triangleright\Delta f.
    \nonumber
%    \label{?d(f(x))}
\end{eqnarray}
In particular:
\begin{equation}
    \boxed{\Delta f=f(x^{*})-f(x)+\delta f(x^{*})}, 
    \label{Df=fx*-fx+dfx*}
\end{equation}
e. g., for $f:x\mapsto y.x$, where `$.$' stands for a bilinear
product,
\begin{equation}
    \delta(y.x)=y.\delta x+\delta y.x^{*}.
    \label{d(xy)=}
\end{equation}

I assume that the source system remains of the form
(\ref{0=LP+Q}, \ref{?R}), hence the `law invariance' statement
(on the operator $U$, not its value!):
\begin{equation}
    \boxed{\delta U=0}, 
    \label{dU=0}
\end{equation} 
and, consequently, $\delta(\underline{U}, \underline{z})=0$.  On the
contrary, (the function) $T$ may vary by (the smooth function) $\delta
T$.  Control is linear if and only if the second derivatives vanish:
\begin{equation}
    0=T''=\delta T''.
    \label{?linear}
\end{equation}
Definition: control is \emph{remote} (from excitation) if and only if
\begin{equation}
    \delta T'=0.
    \label{?remote}
\end{equation}

A perturbation of the unconstrained system $U(T(z))$ is, from 
(\ref{?f(x)*}, \ref{dU=0}), 
\begin{eqnarray}
    U(T(z))^{*}=U(T(z)+\Delta T)&\triangleright&U^{*}, 
%    \label{U*=}
    \nonumber\\
    T(z^{*})-T(z)+\delta T(z^{*})&\triangleright&\Delta T.  
    \label{DT=}
\end{eqnarray}
Taking advantage of (\ref{scaling2}), the perturbed constrained system 
$\underline{U}(T^{*},R_{0}^{*})$ is rescaled, according to the gauge 
transform
\begin{equation}    
    (Q^{\dag *},R_{0}^{*},\underline{\Phi}^{\dag *})
    \rightarrow\alpha
    (Q^{\dag *},R_{0}^{*},\underline{\Phi}^{\dag *}), 
    \alpha=\frac{R_{0}}{R_{0}^{*}},
    \label{R0*->R0}
\end{equation}
so that
\begin{equation}
    R_{0}=R_{0}^{*}.
    \label{R0=R0*}
\end{equation}
Whenever the gauge output is constrained, it is taken as constant,
without loss of generality, which allows to drop the independent
variable $R_{0}$.

From (\ref{Df=fx*-fx+dfx*}), for all operator $u$ on $T(z)$,
\begin{eqnarray}
    \forall(u,\delta u=0),\underline{\Delta u}
    &=&u(\underline{T}^{*})-u(\underline{T})
    \triangleright\underline{u}^{*}-\underline{u}
    \triangleright\delta\underline{u}, \nonumber\\
    \_\Delta&=&\delta\_:
    \label{_D=d_}
\end{eqnarray}
the result of constraint and perturbation on $u(T(z))$ does not depend 
on the order of these operations.  In particular, with 
$u\in\{\id,U\}$, considering (\ref{dU=0}), $\underline{\Delta 
T}=\delta\underline{T}, \underline{\Delta U}=\delta\underline{U}$.  I 
will also often use $\underline{\delta T}$, and one must be careful 
that
\begin{remark}
    $\underline{\delta T}\neq\delta\underline{T}=\underline{\Delta T}$.
\end{remark}

\subsection{Perturbative expressions of the gauge output and the fluxes}

Using (\ref{d(xy)=}) on (\ref{0=LP+Q}, \ref{?R}), and noticing that 
$U$ depends on $T$, that depends on $z$,
\begin{eqnarray}
    \Delta R&=&\<\Delta Q^{\dag}|\Phi^{*}\>+\<Q^{\dag}|\Delta\Phi\>, 
    \label{DR=<DQ, P*>+<Q, DP>}\\
    -L\Delta\Phi&=&\Delta L\Phi^{*}+\Delta Q.
    \label{0=LDP+DLP*+DQ}
\end{eqnarray}
Multiplying (\ref{0=LDP+DLP*+DQ}) by $\<\Phi^{\dag}|$ and using 
(\ref{(0=LP+Q)^+}), 
\begin{equation}
    \<Q^{\dag}|\Delta\Phi\>=\<\Phi^{\dag}|\Delta L\Phi^{*}+\Delta Q\>.
    \label{<Q, DP>=<P, DLP*+DQ>}
\end{equation}
Combining (\ref{DR=<DQ, P*>+<Q, DP>}, \ref{<Q, DP>=<P, DLP*+DQ>}), 
\begin{equation}
    \Delta R=dR_{T}(\Delta T)
    +\<\Phi^{\dag}|\Delta L\Delta\Phi\>+\<\Delta Q^{\dag}|\Delta\Phi\>, 
    \label{DR=2}
\end{equation}
where $dR$ is the differential of $R(T)$:
\begin{equation}
    dR_{T}(dT)=\<\Phi^{\dag}|dL\Phi+dQ\>+\<dQ^{\dag}|\Phi\>.
    \label{d/R=}
\end{equation}

The harmonic part of $\Delta\Phi$ is obtained by projecting 
(\ref{0=LDP+DLP*+DQ}) on the harmonic subspace and inverting:
\begin{equation}
    \Delta\tilde{\Phi}=-\tilde{L}^{-1}(\Delta\tilde{L}\Phi+\Delta\tilde{Q}) 
    -\tilde{L}^{-1}\Delta\tilde{L}\Delta\Phi.
    \label{DTiP=}
\end{equation}
% attention danger ! tu vas l'oublier
I abbreviate $\varphi_{c}\triangleright\varphi$ (and adjointly).  The 
fundamental amplitude $\<\varphi^{\dag}|\Delta\Phi\>$ is determined, 
independently of the fundamental eigenvalue, from (\ref{DR=<DQ, 
P*>+<Q, DP>}, \ref{closurep}):
\begin{eqnarray}
    \Delta R&=&\<Q^{\dag}|\Delta\Phi\>+\<\Delta Q^{\dag}|\Phi\>
    +\<\Delta Q^{\dag}|\Delta\Phi\>, 
    \nonumber
\\
    \<Q^{\dag}|\Delta\Phi\>&=&\<Q^{\dag}|\varphi\varphi^{\dag}|\Delta\Phi\> 
    +\<Q^{\dag}|\Delta\tilde{\Phi}\>.
    \nonumber
\end{eqnarray}
From (\ref{DTiP=}, \ref{algTi}, \ref{d/R=}) and after rearranging, 
\begin{eqnarray}
    \Delta R&=&dR_{T}(\Delta\tilde{T})
    +\<Q^{\dag}|\varphi\varphi^{\dag}|\Delta\Phi\>
    +\<\Phi^{\dag}|\Delta\tilde{L}\Delta\Phi\>
    +\<\Delta Q^{\dag}|\Delta\Phi\>, 
    \nonumber\\
    \tilde{T}&\equiv&(\tilde{L},\tilde{Q},Q^{\dag}).
    \label{-<Q,pp,dP>=}
\end{eqnarray}
\begin{remark}
  $\tilde{T}\neq (\tilde{L},\tilde{Q},\widetilde{Q^{\dag}})$.
\end{remark}

The constraint is applied to (\ref{DR=2}, \ref{DTiP=}, 
\ref{-<Q,pp,dP>=}).  In particular, from (\ref{dU=0}, \ref{_D=d_}, 
\ref{R0=R0*}),
\begin{equation}
    0=\underline{\Delta R}=\delta\underline{R}.
    \label{DR_=0}
\end{equation}
\begin{eqnarray}
    0&=&dR_{\underline{T}}(\delta\underline{T})
    +\<\underline{\Phi}^{\dag}|\delta\underline{L}\delta\underline{\Phi}\> 
    +\<\delta\underline{Q}^{\dag}|\delta\underline{\Phi}\>, 
    \label{0=DR_}
\\
    -\underline{\tilde{L}}\delta\underline{\tilde{\Phi}}
    &=&\delta\underline{\tilde{L}}\ \underline{\Phi}
    +\delta\underline{\tilde{Q}}
    +\delta\underline{\tilde{L}}\delta\underline{\Phi}, 
    \label{DTiP=1}
\\
    -\<\underline{Q}^{\dag}|\underline{\varphi\varphi}^{\dag}|
    \delta\underline{\Phi}\>&=&dR_{\underline{T}}(\delta\underline{\tilde{T}})
    +\<\underline{\Phi}^{\dag}|
    \delta\underline{\tilde L}\delta\underline{\Phi}\>
    +\<\delta\underline{Q}^{\dag}|\delta\underline{\Phi}\>.
    \label{-<Q,pp,dP>=1}
\end{eqnarray}
In agreement with (\ref{dP/dQ^+_}), not only the direct source, as 
expected from (\ref{0=LP+Q}), but also the adjoint source, affects the 
flux, by ways of (\ref{R0=R0*}).  The similarity between the r.  h. s.  
of (\ref{0=DR_}, \ref{-<Q,pp,dP>=1}) helps calculations.

(\ref{0=DR_}, \ref{DTiP=1}, \ref{-<Q,pp,dP>=1}) do not yield perturbed
quantities in a closed form; but, with $\delta
\underline{T}=\mathcal{O}(\epsilon),\epsilon\rightarrow 0$, they yield
$\delta\underline{\Phi}$ up to $\mathcal{O}(\epsilon^{2})$: they are
`perturbative' \cite{Greenspan1}, which allows to solve the
perturbation problem by perturbation theory
\cite{WeinbergWigner,Bender}, as follows.

\subsection{Perturbation series of the fluxes and the control 
variable\label{fluperser}}

Perturbation theory transforms $f(x)^{*}$ into a function of a 
perturbation variable $\epsilon$, according to
\begin{equation}
    \delta x\rightarrow\epsilon\delta x.
    \label{d->ed}
\end{equation}
This is applied to the constrained perturbed system.  The response is 
sought as a power series of $\epsilon$:
\begin{equation}
    (\underline{U},\underline{z})(T+\epsilon\delta T)
    \triangleright(\underline{U},\underline{z})^{*}=
    \sum_{n=0}^{\infty}
    (\underline{U},\underline{z})_{n}(T,\delta T)\epsilon^{n}.
    \label{(U_,z_)*=sum}
\end{equation}
By definition, $(\underline{U},\underline{z})_{n}(T,.)$ is
homogeneous:
\begin{equation*}
    \forall\alpha\in\mathbb{R}^{*}, 
    (\underline{U},\underline{z})_{n}(T,\alpha\delta T) 
    =\alpha^{n}(\underline{U},\underline{z})_{n}(T,\delta T).
%    \label{homog}
\end{equation*}    
I do not study the convergence of the perturbation series: I only find
`formal' power series solutions.

Applying (\ref{Df=fx*-fx+dfx*}, \ref{d->ed}, \ref{DT=}) on $T(z)$,
\begin{equation}
    \underline{T}^{*}=T(\underline{z}^{*})+
    \epsilon\delta T(\underline{z}^{*}).
    \label{DT_=1}
\end{equation}

I assume that $(T,\delta T)(z)$ have power series at
$\underline{z}(T)$.  (\ref{Xoz=sum}, \ref{Xozn=znX'+Xn}) are used to
develop in powers of $\epsilon$ the functions of $\underline{z}^{*}$
in the r.  h. s.  of (\ref{DT_=1})
\begin{equation}    
    \underline{T}^{*}=
    \sum_{n=0}^{\infty}(\underline{z}_{n}\underline{T}'+
    \underline{z}_{n-1}\underline{\delta T}'+
    \underline{T}_{n}(\underline{z}_{1}, \ldots\underline{z}_{n-1})+
    \underline{\delta T}_{n-1}
    (\underline{z}_{1}, \ldots\underline{z}_{n-2}))\epsilon^{n}, 
    \label{T_=sum}
\end{equation}
where the term $n=0$ is correct from (\ref{?X0}) and a variable shift 
on $T$, so that 
$0=\underline{z}_{0}=\underline{z}_{-1}=\underline{\delta T}_{-1}$.

The power series (\ref{(U_,z_)*=sum}, \ref{T_=sum}) are introduced 
into (\ref{0=DR_}, \ref{DTiP=1}, \ref{-<Q,pp,dP>=1}).  By 
convention, $\sum_{\emptyset}=0$, and I define, for all endomorphism 
$D$ on $T$, and also for $D=\delta$,
\begin{eqnarray}
    \forall p\ge 0,\<DT\>_{0p}&\equiv&
    \<\underline{DL^{\dag}\Phi^{\dag}+DQ^{\dag}}|\underline{\Phi}_{p}\>,
    \nonumber\\
    \<DT\>_{p0}&\equiv&
    \<\underline{\Phi}^{\dag}_{p}|\underline{DL\Phi+DQ}\>,
    \nonumber\\
    \<DT\>_{0p+p0}&\equiv&\<DT\>_{0p}+\<DT\>_{p0},
    \nonumber\\
    \<DT\>&\equiv&\<DT\>_{00}=dR_{\underline{T}}(\underline{DT}), 
    \label{?<fT>}\\
    \forall p,q\ge 1,\<DT\>_{pq}&\equiv&
    \<\underline{\Phi}^{\dag}_{p}|\underline{DL\Phi}_{q}\>.
    \label{?<DT>pq}
\end{eqnarray}
By identification at order $\epsilon^{n}$:
\begin{multline}
    \forall n\ge 1,-\underline{z}_{n}\<T'\>=
    \underline{z}_{n-1}\<\delta T'\>+\<T_{n}+\delta T_{n-1}\>+\\ 
    \sum_{1\le p\le n-1}(\underline{z}_{n-p}\<T'\>_{0p}+
    \underline{z}_{n-p-1}\<\delta T'\>_{0p}+
    \<T_{n-p}+\delta T_{n-p-1}\>_{0p}),
    \label{z_n=n-1}
\end{multline}
\begin{multline}
    \forall n\ge 1,-\underline{\tilde{L}}\ \underline{\tilde{\Phi}}_{n}=
    \underline{z}_{n}
    (\underline{\tilde{L}}'\underline{\Phi}+\underline{\tilde{Q}}')+
    \underline{z}_{n-1}(\underline{\delta\tilde{L}}'\underline{\Phi}+
    \underline{\delta\tilde{Q}}')+
    (\underline{\tilde{L}}_{n}+\underline{\delta\tilde{L}}_{n-1})
    \underline{\Phi}+
    \underline{\tilde{Q}}_{n}+\underline{\delta\tilde{Q}}_{n-1}+\\
    \sum_{1\le p\le n-1}
    (\underline{z}_{n-p}\underline{\tilde{L}}'+
    \underline{z}_{n-p-1}\underline{\delta\tilde{L}}'+
    \underline{\tilde{L}}_{n-p}+\underline{\delta\tilde{L}}_{n-p-1})
    \underline{\Phi}_{p},
    \label{TiPn=n}
\end{multline}
\begin{multline}
    \forall n\ge 1,
    -\<\underline{Q}^{\dag}|\underline{\varphi\varphi}^{\dag}|
    \underline{\Phi}_{n}\>=
    \underline{z}_{n}\<\tilde{T}'\>+\underline{z}_{n-1}\<\delta\tilde{T}'\>+
    \<\tilde{T}_{n}+\delta\tilde{T}_{n-1}\>+\\
    \sum_{1\le p\le n-1}(\underline{z}_{n-p}\<\tilde{T}'\>_{0p}+
    \underline{z}_{n-p-1}\<\delta\tilde{T}'\>_{0p}+
    \<\tilde{T}_{n-p}+\delta\tilde{T}_{n-p-1}\>_{0p}).
    \label{-<Q,pp,Pn>=n}
\end{multline}

Furthermore, using (\ref{z_n=n-1}), $\underline{z}_{n}$ is 
eliminated from (\ref{TiPn=n}, \ref{-<Q,pp,Pn>=n}) to 
obtain \emph{recursion relations}. After some rearrangements:
\begin{multline}
    \forall n\ge 1, 
    -\underline{\tilde{L}}\ \underline{\tilde{\Phi}}_{n}=
    \underline{z}_{n-1}(-\frac{\<\delta T'\>}{\<T'\>} 
    (\underline{\tilde{L}}'\underline{\Phi}+\underline{\tilde{Q}}')+
    \underline{\delta\tilde{L}}'\underline{\Phi}+
    \underline{\delta\tilde{Q}}')-\\
    \frac{\<T_{n}+\delta T_{n-1}\>}{\<T'\>}
    (\underline{\tilde{L}}'\underline{\Phi}+\underline{\tilde{Q}}')+
    (\underline{\tilde{L}}_{n}+
    \underline{\delta\tilde{L}}_{n-1})\underline{\Phi}+
    \underline{\tilde{Q}}_{n}+\underline{\delta\tilde{Q}}_{n-1}+\\
    \sum_{1\le p\le n-1}(\underline{z}_{n-p}(-\frac{\<T'\>_{0p}}{\<T'\>}
    (\underline{\tilde{L}}'\underline{\Phi}+\underline{\tilde{Q}}')+
    \underline{\tilde{L}}'\underline{\Phi}_{p})+
    \underline{z}_{n-p-1}(-\frac{\<\delta T'\>_{0p}}{\<T'\>}
    (\underline{\tilde{L}}'\underline{\Phi}+\underline{\tilde{Q}}')+
    \underline{\delta\tilde{L}}'\underline{\Phi}_{p})\\
    -\frac{\<T_{n-p}+\delta T_{n-p-1}\>_{0p}}{\<T'\>}
    (\underline{\tilde{L}}'\underline{\Phi}+\underline{\tilde{Q}}')+
    (\underline{\tilde{L}}_{n-p}+
    \underline{\delta\tilde{L}}_{n-p-1})\underline{\Phi}_{p}), 
    \label{TiPn=n-1}
\end{multline}
% presentation pose pb si dŽnominateurs nuls
% \begin{multline}
%     \forall n\ge 1, 
%     -\<\underline{Q}^{\dag}|\underline{\varphi\varphi}^{\dag}|
%     \underline{\Phi}_{n}\>\\
%     =\underline{z}_{n-1}\<\delta T'\>
%     (\frac{\<\delta\tilde{T}'\>}{\<\delta T'\>}
%     -\frac{\<\tilde{T}'\>}{\<T'\>})
%     +\<T_{n}+\delta T_{n-1}\>
%     (\frac{\<\tilde{T}_{n}+\delta\tilde{T}_{n-1}\>}
%     {\<T_{n}+\delta T_{n-1}\>}
%     -\frac{\<\tilde{T}'\>}{\<T'\>})\\
% %
%     +\sum_{1\le p\le n-1}( 
% %
%     \underline{z}_{n-p}\<T'\>_{0p}
%     (\frac{\<\tilde{T}'\>_{0p}}{\<T'\>_{0p}}
%     -\frac{\<\tilde{T}'\>}{\<T'\>})
% %       
%     +\underline{z}_{n-p-1}\<\delta T'\>_{0p}
%     (\frac{\<\delta\tilde{T}'\>_{0p}}{\<\delta T'\>_{0p}}
%     -\frac{\<\tilde{T}'\>}{\<T'\>})\\
% %          
%     +\<T_{n-p}+\delta T_{n-p-1}\>_{0p}
%     (\frac{\<\tilde{T}_{n-p}+\delta\tilde{T}_{n-p-1}\>_{0p}}
%     {\<T_{n-p}+\delta T_{n-p-1}\>_{0p}}
%     -\frac{\<\tilde{T}'\>}{\<T'\>})
%     ).
%     \label{-<Q,pp,Pn>=n-1}
% \end{multline}
\begin{multline}
    \forall n\ge 1, 
    -\<\underline{Q}^{\dag}|\underline{\varphi\varphi}^{\dag}|
    \underline{\Phi}_{n}\>=\\
    \underline{z}_{n-1}
    (\<\delta\tilde{T}'\>-\<\delta T'\>\frac{\<\tilde{T}'\>}{\<T'\>})+
    (\<\tilde{T}_{n}+\delta\tilde{T}_{n-1}\>-
    \<T_{n}+\delta T_{n-1}\>\frac{\<\tilde{T}'\>}{\<T'\>})+\\
    \sum_{1\le p\le n-1}( 
    \underline{z}_{n-p}
    (\<\tilde{T}'\>_{0p}-\<T'\>_{0p}\frac{\<\tilde{T}'\>}{\<T'\>})+
    \underline{z}_{n-p-1}(\<\delta\tilde{T}'\>_{0p}-
    \<\delta T'\>_{0p}\frac{\<\tilde{T}'\>}{\<T'\>})+\\
    (\<\tilde{T}_{n-p}+\delta\tilde{T}_{n-p-1}\>_{0p}-
    \<T_{n-p}+\delta T_{n-p-1}\>_{0p}\frac{\<\tilde{T}'\>}{\<T'\>})
    ).
    \label{-<Q,pp,Pn>=n-1}
\end{multline}

Finally, the coefficients $(\underline{U},\underline{z})_{n}$ can be 
constructed, up to any order, from the reference flux-to-source 
operator, source and constraint, and their exciting perturbations, 
with the recursion relations (\ref{z_n=n-1}, \ref{TiPn=n-1}, 
\ref{-<Q,pp,Pn>=n-1}), and adjoint relations.  First and second order 
results are
\begin{equation}
    \underline{z}_{1}=-\frac{\<\delta T\>}{\<T'\>}, 
    \label{z_1=}
\end{equation}
\begin{equation}
    \underline{\Phi}_{1}=
    -\frac{1}{\<\underline{Q}^{\dag}|\underline{\varphi}\>}
    (\<\delta\tilde{T}\>-
    \<\delta T\>\frac{\<\tilde{T}'\>}{\<T'\>})\underline{\varphi}-
    \underline{\tilde{L}}^{-1}(-\frac{\<\delta T\>}{\<T'\>} 
    (\underline{\tilde{L}}'\underline{\Phi}+\underline{\tilde{Q}}')+ 
    \underline{\delta\tilde{L}}\ \underline{\Phi}+
    \underline{\delta\tilde{Q}});
    \label{P1=}
\end{equation}
\begin{equation}
    \boxed{
    -\underline{z}_{2}\<T'\>=\underline{z}_{1}(\<\delta T'\>+\<T'\>_{01})+
    \<T_{2}\>+\<\delta T\>_{01}}.
    \label{z_2=}
\end{equation}

In (\ref{z_2=}) appear the various contributions to control variable
response non-linearity: $\<T_{2}\>,\<\delta T'\>$ represent,
respectively, the intrinsic control non-linearity and the cross effect
of control and excitation, vanishing if control is, respectively,
linear (\ref{?linear}) or remote (\ref{?remote}). 
$\<T'\>_{01},\<\delta T\>_{01}$ represent control and exciting
perturbation shielding, as they involve the flux shielding coefficient
$\underline{\Phi}_{1}$, a global consequence of both exciting and
control perturbations.
\begin{remark}
    Because of shielding, the control variable response
    ($\epsilon\mapsto\underline{z}$) may be non-linear,
    notwithstanding linear and remote control.
\end{remark}

\begin{multline*}
    -\underline{\tilde{L}}\ \underline{\tilde{\Phi}}_{2}=
    \underline{z}_{1}(-\frac{\<\delta T'\>+\<T'\>_{01}}{\<T'\>} 
    (\underline{\tilde{L}}'\underline{\Phi}+\underline{\tilde{Q}}')+
    \underline{\delta\tilde{L}}'\underline{\Phi}+
    \underline{\tilde{L}}'\underline{\Phi}_{1}+
    \underline{\delta\tilde{Q}}')-\\
    \frac{\<T_{2}\>+\<\delta T\>_{01}}{\<T'\>}
    (\underline{\tilde{L}}'\underline{\Phi}+\underline{\tilde{Q}}')+
    \underline{\tilde{L}}_{2}\underline{\Phi}+
    \underline{\delta\tilde{L}}\ \underline{\Phi}_{1}+
    \underline{\tilde{Q}}_{2}, 
%    \label{TiP2=}
\end{multline*}
\begin{multline*}
    -\<\underline{Q}^{\dag}|\underline{\varphi\varphi}^{\dag}|
    \underline{\Phi}_{2}\>=\underline{z}_{1}
    (
    (\<\delta\tilde{T}'\>-\<\delta T'\>\frac{\<\tilde{T}'\>}{\<T'\>})+
    (\<\tilde{T}'\>_{01}-\<T'\>_{01}\frac{\<\tilde{T}'\>}{\<T'\>}))+\\
    (\<\tilde{T}_{2}\>-\<T_{2}\>\frac{\<\tilde{T}'\>}{\<T'\>})+
    (\<\delta\tilde{T}\>_{01}-
    \<\delta T\>_{01}\frac{\<\tilde{T}'\>}{\<T'\>}).
%    \label{-<Q, pp, P2>=}
\end{multline*}

For linear control, using (\ref{X''=0}), (\ref{z_n=n-1}, 
\ref{TiPn=n-1}, \ref{-<Q,pp,Pn>=n-1}) become
\begin{multline}
    \forall n\ge 1,-\underline{z}_{n}\<T'\>=
    \underline{z}_{n-1}\<\delta T'\>+
    \delta_{n,1}\<\delta T\>+\<\delta T\>_{0,n-1}+\\
    \sum_{1\le p\le n-1}(\underline{z}_{n-p}\<T'\>_{0p}+
    \underline{z}_{n-p-1}\<\delta T'\>_{0p}),
    \label{z_nlin=n-1}
\end{multline}
\begin{multline}
    \forall n\ge 1,-\underline{\tilde{L}}\ \underline{\tilde{\Phi}}_{n}=
    \underline{z}_{n-1}(-\frac{\<\delta T'\>}{\<T'\>} 
    (\underline{\tilde{L}}'\underline{\Phi}+\underline{\tilde{Q}}')+
    \underline{\delta\tilde{L}}'\underline{\Phi}+
    \underline{\delta\tilde{Q}}')+\\
    \delta_{n,1}(-\frac{\<\delta T\>}{\<T'\>}
    (\underline{\tilde{L}}'\underline{\Phi}+\underline{\tilde{Q}}')+
    \underline{\delta\tilde{L}}\ \underline{\Phi}+
    \underline{\delta\tilde{Q}})-
    \frac{\<\delta T\>_{0,n-1}}{\<T'\>}
    (\underline{\tilde{L}}'\underline{\Phi}+\underline{\tilde{Q}}')+
    \underline{\delta\tilde{L}}\ \underline{\Phi}_{n-1}+\\
    \sum_{1\le p\le n-1}(\underline{z}_{n-p}
    (-\frac{\<T'\>_{0p}}{\<T'\>}
    (\underline{\tilde{L}}'\underline{\Phi}+\underline{\tilde{Q}}')+
    \underline{\tilde{L}}'\underline{\Phi}_{p})+
    \underline{z}_{n-p-1}(-\frac{\<\delta T'\>_{0p}}{\<T'\>}
    (\underline{\tilde{L}}'\underline{\Phi}+\underline{\tilde{Q}}')+
    \underline{\delta\tilde{L}}'\underline{\Phi}_{p})),
    \label{TiPnlin=n-1}
\end{multline}
\begin{multline}
    \forall n\ge 1,
    -\<\underline{Q}^{\dag}|\underline{\varphi\varphi}^{\dag}|
    \underline{\Phi}_{n}\>=\underline{z}_{n-1}
    (\<\delta\tilde{T}'\>-\<\delta T'\>\frac{\<\tilde{T}'\>}{\<T'\>})+\\
    \delta_{n,1}
    (\<\delta\tilde{T}\>-\<\delta T\>\frac{\<\tilde{T}'\>}{\<T'\>})+
    (\<\delta\tilde{T}\>_{0,n-1}-
    \<\delta T\>_{0,n-1}\frac{\<\tilde{T}'\>}{\<T'\>})+\\
    \sum_{1\le p\le n-1}(\underline{z}_{n-p}
    (\<\tilde{T}'\>_{0p}-\<T'\>_{0p}\frac{\<\tilde{T}'\>}{\<T'\>})+
    \underline{z}_{n-p-1}
    (\<\delta\tilde{T}'\>_{0p}-
    \<\delta T'\>_{0p}\frac{\<\tilde{T}'\>}{\<T'\>})).
    \label{-<Q,pp,Pn>lin=n-1}
\end{multline}

\subsection{Perturbation series of a bilinear form of the fluxes}

I seek the power series of $\<DT\>^{*}$, where $D$ is any endomorphism 
on $T$, like $D=d/dz$, such that $\delta D=0$.
\begin{remark}
By (\ref{d/R=}, \ref{?<fT>}), the operator $\<D.\>$ depends implicitly
on $T$, through the fluxes, so that, in general $ \delta(\<DT\>,
\delta\<D.\>, \delta T) \triangleright\Delta\<DT\>\neq\<\delta DT\> $.
\end{remark}
I evaluate $\Delta\<DT\>$ by applying twice the formula for the
perturbation of a bilinear product (\ref{d(xy)=}); after rearranging,
I obtain the perturbative expression
\begin{multline}
    \Delta\<DT\>-\<\delta DT\>=\\
    \<\underline{DL}^{\dag *}\underline{\Phi}^{\dag}+
    \underline{DQ}^{\dag *}|\delta\underline{\Phi}\>+
    \<\delta\underline{\Phi}^{\dag}| 
    \underline{DL}^{*}\underline{\Phi}+\underline{DQ}^{*}\>+
    \<\delta\underline{\Phi}^{\dag}|\underline{DL}^{*}
    \delta\underline{\Phi}\>.
    \label{D<ft>=}
\end{multline}

I assume that $D$ is linear.  Applying $D$ and (\ref{d->ed}) on
(\ref{DT=}), noticing that $D$ commutes, by linearity, with $\delta$
and $'$, I obtain that (\ref{T_=sum}), where $T\rightarrow DT$, still
holds, giving the power series of $\underline{DT}^{*}$.  This power
series, and that of $(\underline{U}, \underline{z})^{*}$
(\ref{(U_,z_)*=sum}), are inserted into (\ref{D<ft>=}), yielding the
power series
\begin{equation}
  \delta(\<DT\>, \delta\<D.\>, \epsilon\delta T)\triangleright
  \Delta\<DT\>\equiv
  \sum_{n=1}^{\infty}\<DT\>_{n}(T, \delta T)\epsilon^{n}, 
  \label{d<fT>=}
\end{equation}
the coefficients of which are determined by identification. 
Eventually,
\begin{equation}
  (\<DT\>, \delta\<D.\>, \epsilon\delta T)^{*}\triangleright
  \<DT\>^{*}\equiv
  \sum_{n=0}^{\infty}\<DT\>_{n}(T, \delta T)\epsilon^{n}, 
  \label{<fT>*=}
\end{equation}
\begin{multline}
    \forall n\ge 0, \<DT\>_{n}(T, \delta T)=\underline{z}_{n}\<DT'\>+
    \underline{z}_{n-1}\<D\delta T'\>+\<DT_{n}+D\delta T_{n-1}\>+\\
    \sum_{1\le p\le n}(
    \underline{z}_{n-p}\<DT'\>_{0p+p0}+
    \underline{z}_{n-p-1}\<D\delta T'\>_{0p+p0}+
    \<DT_{n-p}+D\delta T_{n-p-1}\>_{0p+p0})+\\
    \sum_{\substack{1\le p_{1}\le n-1\\1\le p_{2}\le n-p_{1}}}
    (\underline{z}_{n-p_{1}-p_{2}}\<DT'\>_{p_{1}p_{2}}+
    \underline{z}_{n-p_{1}-p_{2}-1}\<D\delta T'\>_{p_{1}p_{2}}+
    \<DT_{n-p_{1}-p_{2}}+D\delta T_{n-p_{1}-p_{2}-1}\>_{p_{1}p_{2}}).
    \label{<fT>n=}
\end{multline}
%% 
 % Moreover, $\<DT\>_{n}$ is homogeneous:
 % \begin{equation*}
 %     \forall\alpha\in\mathbb{R}^{*},\<DT\>_{n}(T,\alpha\delta T)
 %     =\alpha^{n}\<DT\>_{n}(T,\delta T).
 % %    \label{homog1}
 % \end{equation*}
 %%

In (\ref{<fT>n=}), $\underline{z}_{n}$ is known and need not be 
eliminated, which allows to group terms nicely:
\begin{multline}
    \forall n\ge 0, \<DT\>_{n}(T, \delta T)=\\
    \sum_{\substack{0\le p_{1}, p_{2}\le n\\p_{1}+p_{2}\le n}}
    (\underline{z}_{n-p_{1}-p_{2}}\<DT'\>_{p_{1}p_{2}}+
    \underline{z}_{n-p_{1}-p_{2}-1}\<D\delta T'\>_{p_{1}p_{2}}+
    \<DT_{n-p_{1}-p_{2}}+D\delta T_{n-p_{1}-p_{2}-1}\>_{p_{1}p_{2}}).
    \label{<fT>n=1}
\end{multline}

First and second order results are
\begin{equation*}
    \<DT\>_{1}=\underline{z}_{1}\<DT'\>+\<D\delta T\>+\<DT\>_{01+10}, 
%    \label{<fT>1=}
\end{equation*}
\begin{multline*}
    \<DT\>_{2}=\underline{z}_{2}\<DT'\>+\<DT_{2}\>+
    \underline{z}_{1}(\<D\delta T'\>+\<DT'\>_{01+10})+\\
    \<D\delta T\>_{01+10}+\<DT\>_{02+20}+\<DT\>_{11}.
    \label{<fT>2=}
\end{multline*}

\section{Weight definition and observability\label{weight}}

\subsection{The balance equation and weight definition \label{baleq}}

For an infinitesimal perturbation of the constrained system,
(\ref{0=DR_}) becomes:
\begin{equation}
    0=dR_{\underline{T}}(d\underline{T}).
    \label{0=d/R(d/T)}
\end{equation}
Applying (\ref{Df=fx*-fx+dfx*}) to $d(T(z))$, then the constraint, 
with (\ref{_D=d_}):
\begin{equation}
    d(T(z))=T'(z)dz+dT(z);
    d\underline{T}=\underline{T}'d\underline{z}+\underline{dT}.
    \label{d/T=T'dz+d/T_}
\end{equation}
I introduce (\ref{d/T=T'dz+d/T_}) into (\ref{0=d/R(d/T)}), use 
(\ref{?<fT>}), where $D\rightarrow d$, to obtain the \emph{balance} 
equation
\begin{equation}
    0=\<T'\>d\underline{z}+\<dT\>.
    \label{balance1}
\end{equation}
$\<T'\>$ is the `differential weight' of the exciting variable $z$. 
(\ref{balance1}) is indeed equivalent to (\ref{z_1=}).

From the chain rule, the differential weight is just the derivative of 
the unconstrained gauge output (at constant $T$), with respect to $z$, 
hence an observable:
\begin{equation}
  \boxed{\<T'\>=(R\circ T)'(\underline{z})}.
  \label{difwobs}
\end{equation}

I show up the exciting variable $z_{1}$, driving $\delta T$, while
the original control variable is noted $z_{2}$:
\begin{equation}
    \exists T_{2},(T+\delta T)(z_{2})=T_{2}(z_{1},z_{2}).
    \label{?T2}
\end{equation}
For a given exciting perturbation $\delta T$, (the function) $T_{2}$
remains constant.
\begin{remark}
  An experiment that is not reproducible is still represented by a
  constant $T_{2}$, taking for $z_{1}$ the quasi-stationary time.
\end{remark}

I assume for $T_{2}(.,z_{2})$ the same regularity properties as for
$T_{2}(z_{1},.)$.  Thus, the variables $(z_{1},z_{2})$ are
exchangeable, and either may serve as the control variable.  I define
an exchange operator:
\begin{equation}
    ET_{2}(z_{2},z_{1})\equiv T_{2}(z_{1},z_{2}).
    \label{?E}
\end{equation}
The constraint operator applies to each partial function 
$T_{2}(.,z_{2}),T_{2}(z_{1},.)$.  For example, with $z_{2}$ as the 
control variable,
\[
    \underline{u}(T_{2}(z_{1},.))=u(T_{2}(z_{1},\underline{z}_{2})),
    \underline{z}_{2}(T_{2}(z_{1},.))\triangleright\underline{z}_{2}.
\]
The functions $z_{2}\mapsto\underline{z}_{1}(T_{2}(.,z_{2}))$ and 
$z_{1}\mapsto\underline{z}_{2}(T_{2}(z_{1},.))$ are inverse to each 
other and the constrained value of $u(T_{2}(z_{1},z_{2}))$ does not 
depend on the choice of the control variable.  For all 
$(z_{1},z_{2})$, such that $R(T_{2}(z_{1},z_{2}))=R_{0}$,
\begin{align}
    (z_{1},\underline{z}_{2})=(z_{1}&,z_{2})=(\underline{z}_{1},z_{2}),
%    \label{(z1_,z2)=(z1,z2_)}
    \nonumber\\
    \underline{u}(T_{2}(z_{1},.))=u(T_{2}(z_{1}&,z_{2}))
    =\underline{u}(T_{2}(.,z_{2})).
%    \label{z1<->z2}
    \nonumber
\end{align}

With (\ref{?T2}), the balance equation (\ref{balance1}) (with $z_{2}$ 
as the control variable) takes the nearly symmetric form
\begin{equation}
    0=\<\partial_{1}T_{2}(.,\underline{z}_{2})\>dz_{1}
    +\<\partial_{2}T_{2}(z_{1},.)\>d\underline{z}_{2}.
    \label{balance2}
\end{equation}
The differential weight of $z_{1}$ (controlled by $z_{2}$) is
\begin{equation}    
    w_{1}(T_{2},z_{1})=\<\partial_{1}T_{2}(z_{1},.)\>,
    \label{?w1}
\end{equation}
and symmetrically, so that (\ref{balance2}) becomes
\begin{equation}
    0=w_{1}(T_{2},z_{1})dz_{1}
    +w_{2}(T_{2},\underline{z}_{2})d\underline{z}_{2}.   
    \label{balance3}
\end{equation}

I define the (integral) weight (of $z_{1}$), 
\begin{equation}
    Z_{1}(T_{2},z_{1})\equiv\int_{0}^{z_{1}}w_{1}(T_{2},z_{1}')dz_{1}'.
    \label{?Z1}
\end{equation}
The function $Z_{1}(T_{2},.)$ is the (non-linear) weight scale.  By
integration of (\ref{balance3}), the sum of weight perturbations, over
both variables, is zero:
\begin{equation}
    Z_{1}(T_{2},z_{1})+
    Z_{2}(T_{2},\underline{z}_{2}(T_{2}(z_{1},.)))
    -Z_{2}(T_{2},\underline{z}_{2}(T_{2}(0,.)))=0.
    \label{Z1+Z2=01}
\end{equation}
(\ref{difwobs}, \ref{Z1+Z2=01}) are the basis of the weighing method,
announced by (\ref{Z1+Z2=0}).

\subsection{Weight perturbation theory\label{WPT}}

From law invariance (\ref{dU=0}) (on functionals):
\begin{equation}
    0=\delta w_{1}=\delta Z_{1}.
    \label{dZ=0}
\end{equation}

A general differential weight perturbation is 
\begin{equation}
    \delta(w_{1}(T_{2}, z_{1}), 0, (\delta T_{2}, \delta z_{1}))
    \triangleright\Delta w_{1}. 
    \label{dw=}
\end{equation}
By definition (\ref{?w1}),
\begin{equation}  
    \Delta w_{1}=
    \delta(\<\partial_{1}T_{2}(z_{1},.)\>, \delta\<\partial_{1}.\>, 
    \delta(T_{2}(z_{1},.)))
    \nonumber
\end{equation}
which is just (\ref{d<fT>=}), where
\begin{equation}
    T\rightarrow T_{2}(z_{1},.);    
    D\rightarrow\partial_{1};
    '\rightarrow\partial_{2};
    \epsilon\rightarrow 1. 
    \label{dw(dz=0)=}
\end{equation}

Using the properties of definite integral in (\ref{?Z1}), an arbitrary
weight perturbation is
\begin{equation}
    \delta(Z_{1}(T_{2}, z_{1}), 0, \delta(T_{2}, z_{1}))=
    \int_{z_{1}}^{z_{1}^{*}}w_{1}(T_{2}, z_{1})dz_{1}'+
    \int_{0}^{z_{1}^{*}}
    \delta(w_{1}(T_{2}, z_{1}'), 0, (\delta T_{2}, 0))dz_{1}'.
    \label{DZ1=}
\end{equation}
The last integrand is just (\ref{dw=}), where \[\delta 
z_{1}\rightarrow 0; z_{1}\rightarrow z_{1}'.\]

(\ref{dw=}) has another important application. I write (\ref{?Z1}) as
\begin{equation}
    Z_{1}(T_{2}, z_{1})=
    \int_{0}^{z_{1}}(w_{1}(T_{2}, 0), 0, (0, z_{1}'))^{*}dz_{1}'=
    w_{1}(T_{2}, 0)z_{1}+\int_{0}^{z_{1}}
    \delta(w_{1}(T_{2}, 0), 0, (0, z_{1})).
    \label{Z=z<T'>}
\end{equation}
The last integrand is just (\ref{dw=}), where \[z_{1}\rightarrow 0;
\delta T_{2}\rightarrow 0; \delta z_{1}\rightarrow z_{1}.\] Therefore,
the weight scale itself is obtained as a perturbation series, which
is, more precisely, (\ref{d<fT>=}), where
\begin{equation*}
    \delta T\rightarrow T_{2}(z_{1}',.)-T_{2}(0,.);
    T\rightarrow T_{2}(0,.); 
    D\rightarrow\partial_{1};
    '\rightarrow\partial_{2};
    \epsilon\rightarrow 1. 
%    \label{dw(dT=0)=}
\end{equation*}

I complete the determination of the weight scale only in the case of a
linear exciting variable: $\partial^{2}_{11}T_{2}=0$, equivalent to
\begin{equation}
    T_{2}(z_{1}, z_{2})=T(z_{2})+z_{1}\delta T(z_{2});
    \delta T(z_{2})\equiv\partial_{1}T_{2}.
    \label{T2lin=}
\end{equation}
With (\ref{T2lin=}), the remote control condition (\ref{?remote})
becomes $\partial^{2}_{12}T_{2}=0$.  In (\ref{Z=z<T'>}), $z_{1}'$ is
used as the perturbation variable:
\begin{equation*}
    (w_{1}(T_{2},0),0,(0,z_{1}'))^{*}=
    (\<\partial_{1}T_{2}\>,\delta\<\partial_{1}.\>,z_{1}'\delta T)^{*}
    \triangleright\<\delta T\>^{*}=
    \sum_{n=0}^{\infty}\<\delta T\>_{n}z_{1}^{\prime n},
%    \label{<fT*>lin=}
\end{equation*}
and the coefficients $\<\delta T\>_{n}$ are obtained from
(\ref{<fT>n=}, \ref{<fT>n=1}), where
\begin{equation*}
    \epsilon\rightarrow z_{1}';DT\rightarrow\delta T;
    D\delta T\rightarrow 0:
%    \label{dw(dT=0)=lin}
\end{equation*}
\begin{multline*}
    \forall n\ge 0,\<\delta T\>_{n}=
    \underline{z}_{n}\<\delta T'\>+\<\delta T_{n}\>+
    \sum_{1\le p\le n}(
    \underline{z}_{n-p}\<\delta T'\>_{0p+p0}+\<\delta T_{n-p}\>_{0p+p0})+\\
    \sum_{\substack{1\le p_{1}\le n-1\\1\le p_{2}\le n-p_{1}}}
    (\underline{z}_{n-p_{1}-p_{2}}\<\delta T'\>_{p_{1}p_{2}}+
    \<\delta T_{n-p_{1}-p_{2}}\>_{p_{1}p_{2}});
%    \label{<fT>nlin=}
\end{multline*}
\begin{equation}
    \forall n\ge 0,\<\delta T\>_{n}
    =\sum_{\substack{0\le p_{1},p_{2}\le n\\p_{1}+p_{2}\le n}}
    (\underline{z}_{n-p_{1}-p_{2}}
    \<\delta T'\>_{p_{1},p_{2}}+\<\delta T_{n-p_{1}-p_{2}}\>_{p_{1}p_{2}}).
    \label{<fT>nlin=1}
\end{equation}
First and second order results are
\begin{eqnarray*}
    \<\delta T\>_{1}&=&\underline{z}_{1}\<\delta T'\>
    +\<\delta T\>_{01+10},
%    \label{<dT>1lin=}
\\
    \<\delta T\>_{2}&=&\underline{z}_{2}\<\delta T'\>
    +\<\delta T_{02}\>+\underline{z}_{1}\<\delta T'\>_{01+10}
    +\<\delta T\>_{02+20}+\<\delta T\>_{11}.
%    \label{<dT>2lin=}
\end{eqnarray*}

Evaluating the integral in (\ref{Z=z<T'>}), and with 
$z_{1}\rightarrow\epsilon$, to remember that the exciting variable is 
linear,
\begin{equation}
  \boxed{Z_{1}(T_{2},\epsilon)=\sum_{n=0}^{\infty}
  \<\delta T\>_{n}\frac{\epsilon^{n+1}}{n+1}}.
  \label{what}
\end{equation}

\subsection{Operator perturbation weighing\label{pertmeas}}

I define an \emph{ideal} instrument by its parameters $T$, consisting
of linear functions of the control variable (there is no interest in
non-linear ideal control); the measured object is a perturbation
$\delta T$, with remote control.  The combination of the instrument
and the measured object is represented by the functional parameters
$T_{2}$,
\begin{equation}
    T_{2}(\epsilon,z)\equiv T(z)+\epsilon\delta T, 
    \underline{z}(T_{2}(0,.))\equiv 0.
    \label{?T2lin}
\end{equation}

Observables are not available for $T_{2}$, because of its ideal part
$T$, but on some approximate realization $T_{2}^{*}$.  $\delta
T_{2}=T_{2}^{*}-T_{2}$ is the realization error, meaning it is 
impossible to realize exactly the ideal experiment.  From
(\ref{Z1+Z2=01}),
\begin{multline}    
    -Z_{1}(T_{2},\epsilon)=
    Z_{2}(T_{2},\underline{z}(T_{2}(\epsilon,.))-Z_{2}(T_{2},0)\approx\\
    Z_{2}(T_{2}^{*},\underline{z}(T_{2}^{*}(\epsilon,.))-
    Z_{2}(T_{2}^{*},0).
    \label{Z1+Z2=02}
\end{multline}
The error analysis is left for the next section.  From (\ref{?Z1},
\ref{difwobs}),
\begin{equation}
    Z_{2}(T_{2}^{*},\underline{z}(T_{2}^{*},\epsilon))-Z_{2}(T_{2}^{*},0) 
    =\int_{0}^{\underline{z}(T_{2}^{*},\epsilon)}
    \partial_{2}R\circ 
    T_{2}^{*}(\underline{\epsilon}(T_{2}^{*},z),z)dz,
    \label{howto}
\end{equation}
which I write, more lightly, by impliciting $T_{2}^{*}$:
\begin{equation}
    \boxed{Z_{2}(\underline{z}(\epsilon))-Z_{2}(0)= 
    \int_{0}^{\underline{z}(\epsilon)}
    \partial_{2}R^{*}(\underline{\epsilon}(z),z)dz}.
    \label{howto1}
\end{equation}
As explained in section \ref{baleq}, the functions
$\underline{\epsilon}$ and $\underline{z}$ (at constant $T_{2}$) are
inverse to each other.

(\ref{howto1}) explains \emph{how to} map observables onto weight, by
universal operations, like summation, independent of the system
parameters.  (\ref{what}) expresses \emph{what is} weight and how to
to compute weight, by universal operations, from the system
parameters, representing the instrument and the measured object.  The
function $Z_{1}(T_{2},.)$ is a power series, bijectively related to
its coefficients $\<\delta T\>_{n}$.  As control in $T_{2}$ is linear
and remote, (\ref{<fT>nlin=1}) becomes
\begin{equation}
    \forall n\ge 0,\<\delta T\>_{n}= 
    \sum_{\substack{0\le p_{1},p_{2}\le n\\p_{1}+p_{2}=n}} 
    \<\delta T\>_{p_{1}p_{2}}.
    \label{<dT>n=}
\end{equation}
In particular, from (\ref{?<fT>}, \ref{d/R=}),
\begin{equation*}
    \<\delta T\>_{00}=\<\delta T\>_{0}=\<\delta T\>=
    \<\Phi^{\dag}|\delta L\Phi\>+
    \<\Phi^{\dag}|\delta Q\>+\<\delta Q^{\dag}|\Phi\>.
%    \label{eq:}
\end{equation*}
There are simplifications also in the recursion relations 
(\ref{z_n=n-1}, \ref{TiPn=n-1}, \ref{-<Q,pp,Pn>=n-1}).

If the sequence of perturbed fluxes $\underline{\Phi}_{n}$ is a basis
(and adjointly) and $0=\delta Q=\delta Q^{\dag}$, then $\delta L$ is
completely determined (except for a scalar factor) by its matrix
elements $\<\delta T\>_{p_{1}p_{2}}$, defined by (\ref{?<DT>pq}). 
From (\ref{<dT>n=}), $\<\delta T\>_{n}$ is the finite sum, invariant
by transposition, of the diagonal $p\mapsto\<\delta T\>_{n-p,p}$.  The
$\<\delta T\>_{n}$ do not fully determine $\delta L$, except, for
example, if $\delta L$ has only one coefficient on each diagonal.  

If $\mathcal{H}$ is not separable (has no basis \cite[\S 3.7, p. 
95]{Vulikh}), then $\delta L$ cannot be obtained from observables.  We
must content ourselves with weight, which does not completely
determine $\delta L$, but can be used to check postulated $\delta L$,
given the perturbed fluxes.  Or, knowing $\delta L$, perturbed flux
calculations can be checked, from measured weights.

\subsection{Error analysis}

From (\ref{Df=fx*-fx+dfx*}), the error on the measured weight scale is
\begin{equation}
    Z_{2}^{*}(T_{2}^{*},.)-Z_{2}(T_{2},.)
    =(Z_{2}(T_{2}^{*},.)-Z_{2}(T_{2},.))+\delta Z_{2}(T_{2}^{*},.).
    \label{dZ2T2}
\end{equation}
The first term in the r.  h. s.  is the realization error, that may be
evaluated, \emph{just like any perturbation}, from weight perturbation
formulae (\ref{DZ1=}, \ref{<fT>*=}, \ref{<fT>n=}), after the variable
exchange $Z_{2}(T_{2},.)=Z_{1}(ET_{2},.)$ (\ref{?E}).

The realization error could be removed by taking $T_{2}\equiv
T_{2}^{*}$, thus attaching the weight definition to a real (non-ideal)
standard (`old-style' metrology).

The second term in the r.  h. s.  of (\ref{dZ2T2}), $\delta
Z_{2}(T_{2}^{*},.)$, is a processing error, meaning that observables
are not processed exactly as demanded by (\ref{howto}), for example,
because of discretization errors.  Processing errors such that
\begin{equation}
    \exists f,Z_{2}^{*}(T_{2},.)=Z_{2}(f(T_{2}),.),
    \label{rederr}
\end{equation}
are reducible, by definition (\ref{rederr}), to realization errors. 
Not all processing errors are reducible.

Errors in the l.  h. s.  of (\ref{Z1+Z2=02}) are mathematically
treated just as errors in the r.  h. s.  (except for variable
exchange).  Although $T$ is ideally defined, one may have to compare
weight scales resulting from different definitions of $T$.  An error
on $Z_{1}$ may also occur in computation, e.  g. a power series
truncation.  From (\ref{dZ=0}), an error on $Z_{1}$, which is not
reducible in the sense of (\ref{rederr}), breaks the law of
flux-to-source operator and constraint linearity.

The weight scale is actually obtained approximately, and only on a
finite set of $\epsilon$.  Something must be said on error
propagation, from the weight scale points to its power series
coefficients.  

Firstly, I consider only the effect of discreteness of $\epsilon$.  I
assume that $Z_{1}(T_{2},\epsilon)$ is exactly known, but only for
$N+2, N\ge 0$ discrete values of $\epsilon$.  An interpolating
polynomial of degree $N+1$ can be constructed, giving (approximately)
the coefficients $\<\delta T\>_{0\le n\le N}$; higher order
coefficients are completely undetermined: there is a cut-off between
known and unknown coefficients, at $n=N$.  If $N=0$ (only two values
of $\epsilon$ are realized), then only $\<\delta T\>_{0}$ is obtained.

Secondly, I take into account, not only discreteness, but also errors
on weight values.  Errors propagate non-uniformly to the coefficients
$\<\delta T\>_{n}$, the error on $\<\delta T\>_{n}$ increasing with
$n$, at given $N$, and decreasing with $N$ at given $n$.  Hence the
interest of taking large $N$, even if only $\<\delta T\>_{0}$ is
sought.  $N\ge 1$ reveals and allows to correct the shielding error on
$\<\delta T\>_{0}$ obtained with $N=0$.

\section{Conclusion}

I applied perturbation theory to a linear source problem, allowing for
a critical limit, the consequences of which were non-linear response,
ill-posedness, a regularizing constraint, and the functional character
of parameters.

The main hypotheses were the constraint linearity and the
existence of a stable, spectrally separated, harmonic hyperplane.  (In
spatially extended systems, spectral separation usually follows from
boundary conditions.)

The basic tool was linear (or multi-linear) vector algebra.  The
perturbation of functional parameters and the constraint application
were delicate and required specific notations.  Progress was achieved
by the thorough use of symmetries (the last three being familiar in
theoretical physics):
\begin{itemize}
    \item duality (`adjointness'),
    
    \item commutation relations (between the constraint operator and
    variational symbols),
    
    \item exchange (of the exciting and control variables),
    
    \item gauge invariance (results are independent of the gauge 
    output unit).
\end{itemize}

Perturbation theory was then used to propose a weight definition and
measurement method.  More precisely, the weight defining functional
$Z_{1}$ and the weighing functional $Z_{2}$, needed in
(\ref{Z1+Z2=0}), were constructed, showing weight as a secondary
observable, except for a realization error, which was analyzed, from
the same perturbation analysis.

The weighing functional is an integro-differential processor (filter)
on both the unconstrained (open loop) and constrained (closed loop)
system responses.

The weight scale $Z_{1}(T_{2},.)$ is a power series, whose
coefficients are (in the simplest case) diagonal sums of the
flux-to-source perturbation matrix $\delta L$.  As this relation is
generally not invertible, $\delta L$ cannot be fully determined.

The present work is not mathematically complete, and does neither
treat in detail any particular application.  Emphasis was put into
discovering and solving a new problem, operator weighing.

The mathematical model and the weighing method may actually apply,
more or less easily, to high-gain feedback linear amplifiers, with any
number of degrees of freedom: electronic amplifiers, photomultipliers,
nuclear fission chain reactors \cite{Foell, Bell} and computational
models thereof.

On the mathematical side, progress may be sought in the formulation of
the abstract problem, the analytic aspects of its solution (the
conditions of convergence of power series), and carrying out large
symbolic calculations.
%The study of power series convergence (except in section \ref{secExa})
%is left for further work.

Systems are often studied with the approximation of linear response,
which fails for non-linear systems, but also, as I pointed out, for a
linear system (especially quasi-critical).  Measurement methods based
on the linear response approximation lack of an essentially non-linear
error analysis, and hardly have any quantitative interest.

% In the nuclear fission chain reactor context, the flux-to-source
% operator is a transport operator \cite[ch.  XXI]{Dautray} and weights
% are neutron integral cross sections, from which differential cross
% sections can be statistically adjusted.  There was an effort, possibly
% of historical importance, in perturbation experiments \cite{Anderson,
% Foell} and theory \cite{Weinberg2, WeinbergWigner}.  Nevertheless, the
% actual link between experiments and theory remained loose, possibly
% because of some confusion between the exactly critical and
% quasi-critical cases, and the abusive use of reactivity (a particular
% form of control variable dependence), inhibiting a theory of
% measurement based on observability and non-linear response.  Thus, the
% analysis of perturbation experiments \cite{Foell, Bell} has remained
% often at the level of the linear approximation, which does not allow
% error analysis, so that it persists as a stable state of ignorance,
% although some works \cite{Hansen, Greenspan1, Greenspan2} contained
% the incentive for a more elaborate treatment.

\appendix

\section{Composition of power series\label{secComPowSer}}

The power series of the compound function $T\circ z$ is sought.  $z$ 
has a power series at zero and $T$ has a Taylor series at 
$z_{0}=z(0)$:
\begin{equation*}
    T(z)=\sum_{n=0}^{\infty}\frac{1}{n!}T^{(n)}(z_{0})(z-z_{0})^{n}.
%    \label{X=sum}
\end{equation*}

The multinomial formula, 
\begin{equation*}
    \forall n\ge 0, 
    (z(\epsilon)-z_{0})^{n}=(\sum_{p=1}^{\infty}z_{p}\epsilon^{p})^{n}
    =\sum_{q_{p}\ge 0, \sum q_{p}=n}n!\prod_{p\ge 1}
    \frac{z_{p}^{q_{p}}}{q_{p}!}\epsilon^{\sum p q_{p}}, 
%    \label{multinom}
\end{equation*}
is multiplied by $T^{(n)}/n!$, summed over $n$, and the terms 
of same power in $\epsilon$ are gathered:
\begin{eqnarray}
    T\circ z(\epsilon)&=&
    \sum_{n=1}^{\infty}(T\circ z)_{n}(z_{0})\epsilon^{n}, 
    \label{Xoz=sum}\\
    (T\circ z)_{n}&=&\sum_{q_{p}\ge 0, \sum p q_{p}=n}
    T^{(\sum q_{p})}\prod_{p\ge 1}\frac{z_{p}^{q_{p}}}{q_{p}!}.
    \nonumber
\end{eqnarray}
The sum in the r. h. s.  is split into a term corresponding to the
sequence $q_{p}=\delta_{p, n}$, and a remainder $T_{n}$ ($T_{1}=0$):
\begin{eqnarray}
    \forall n\ge 1, (T\circ z)_{n}&=&z_{n}T'+T_{n}(z_{1}, \ldots z_{n-1}), 
    \label{Xozn=znX'+Xn}\\
    \forall n\ge 2, T_{n}(z_{1}, \ldots z_{n-1})&\equiv&
    \sum_{p<n, q_{p}\ge 0, \sum p q_{p}=n}
    T^{(\sum q_{p})}\prod_{1\le p<n}\frac{z_{p}^{q_{p}}}{q_{p}!}.
    \label{?Xn}\\
    T_{0}&\equiv&T, 
    \label{?X0}\\
    T_{1}&=&0, 
    \nonumber\\
    T_{2}(z_{1})&=&T''z_{1}^{2}/2, 
    \label{X2=}\\
    T_{3}(z_{1}, z_{2})&=&T''z_{1}z_{2}+T'''z_{1}^{3}/3!.
    \nonumber
\end{eqnarray}

If $T$ is linear (\ref{?linear}), then
\begin{equation}
    \forall n\ge 2, T_{n}=0.    
    \label{X''=0}
\end{equation}
\begin{proof}
If $T$ is linear and $T^{(\sum q_{p})}$ in (\ref{?Xn}) is not zero, then 
\[
    \sum_{1\le p<n} q_{p}=1;\exists!m<n, q_{p}=\delta_{p, m};
    \sum_{1\le p<n} p q_{p}=m=n, 
\] 
which is inconsistent.
\end{proof}

\section{Finite-dimensional examples\label{secExa}}

The reference gauge output $R_{0}$ is set to unity.  

I assume, for simplicity, linear and remote control:
\begin{eqnarray}
    L(z')=A+z'B,\delta B=0,\delta L(z')=C\equiv\delta A,\nonumber\\
    L_{2}(\epsilon,z')=L(z')+\epsilon\delta L(z')=A+z'B+\epsilon C,
    \nonumber
\end{eqnarray}
where $A,B,C$ are constant matrices; $A$ has the eigenvalue zero;
sources are unexcited ($0=\delta Q=\delta Q^{\dag}$) and uncontrolled
($0=Q'=Q^{\dag\prime}$); there is no realization error ($\delta
T_{2}=0$).

I will obtain directly the fluxes, the gauge output, the control
variable and flux response, and check (\ref{Z1+Z2=01}, \ref{what},
\ref{howto1}).  The functional operations, like perturbation and
constraint, will appear concretely.

\subsection{One dimension}

$A=0$, the harmonic subspace is $\{0\}$.
\begin{eqnarray*}
    \Phi^{*}(\epsilon,z')&=&-\frac{Q}{z'B+\epsilon C};\\
    R^{*}(\epsilon,z')&=&-\frac{Q^{\dag}Q}{z'B+\epsilon C},
    1=-\frac{Q^{\dag}Q}{\underline{z}'B+\epsilon C}.
\end{eqnarray*}
With the new variable $z\equiv z'+Q^{\dag}Q/B$,
\begin{eqnarray}
    \underline{z}(\epsilon)&=&-\frac{C\epsilon}{B},\underline{z}(0)=0,
    \nonumber\\
    \Phi^{*}(\epsilon,z)&=&\frac{Q}{Q^{\dag}Q-zB-\epsilon C},
    \nonumber\\
    \underline{\Phi}^{*}(\epsilon)&=&\frac{1}{Q^{\dag}},
    \underline{\Phi}^{\dag *}(\epsilon)=\frac{1}{Q}, 
    \nonumber\\
    R^{*}(\epsilon,z)&=&\frac{Q^{\dag}Q}{Q^{\dag}Q-zB-\epsilon C}.
    \nonumber
\end{eqnarray}
From (\ref{?w1}, \ref{?Z1}, \ref{howto1}):
\begin{eqnarray}
    Z_{1}(\epsilon)&=&
    \underline{\Phi}^{\dag *}C\underline{\Phi}^{*}\epsilon=
    \frac{C\epsilon}{Q^{\dag}Q};
    \nonumber\\   
    \partial_{2}R^{*}(\epsilon,z)&=&
    \frac{BQ^{\dag}Q}{(Q^{\dag}Q-zB-\epsilon C)^{2}},
    \partial_{2}R^{*}(\underline{\epsilon},z)=\frac{B}{Q^{\dag}Q};
    \nonumber\\
    Z_{2}(\underline{z}(\epsilon))-Z_{2}(0)&=&
    \int_{0}^{\underline{z}(\epsilon)}
    \partial_{2}R^{*}(\underline{\epsilon}(z),z)dz=
    \frac{B\underline{z}}{Q^{\dag}Q}=-Z_{1}(\epsilon).
    \nonumber
\end{eqnarray}  
$\delta L/(Q^{\dag}Q)$ is the secondary observable, resulting from the
weighing method.  
%If $Q^{\dag}Q$ is not known, then excitation ratios
%can still be measured.

\subsection{Two dimensions, linear control variable response\label{dim2}}

The complementary matrix (or transposed comatrix) $\overline{A}$ of a 
matrix $A$ is
\begin{equation*}
    \forall A=\begin{pmatrix}a_{11}&a_{12}\\a_{21}&a_{22}&\end{pmatrix},
    \overline{A}
    \equiv\begin{pmatrix}a_{22}&-a_{12}\\-a_{21}&a_{11}&\end{pmatrix}.
%    \label{?ovA}
\end{equation*}
\begin{equation}
    \overline{A}A=A\overline{A}=\det(A)1,
    \overline{\overline{A}}=A,\overline{A}^{\dag}=\overline{A^{\dag}}.
    \label{propovA}
\end{equation}
The operator $A\mapsto\overline{A}$ is linear and the non-linearity of
$A\mapsto A^{-1}=\det(A) ^{-1}\overline{A}$ lays fully in the inverse
determinant.

Let the `codeterminant' $A*B$ of two matrices $A,B$ be
\begin{equation*}
    A*B\equiv
    \begin{vmatrix}
	a_{11}&a_{12}\\
	b_{21}&b_{22}
    \end{vmatrix}
    +
    \begin{vmatrix}
	b_{11}&b_{12}\\
	a_{21}&a_{22}
    \end{vmatrix}
%     =
%     \begin{vmatrix}
% 	a_{11}&b_{12}\\
% 	a_{21}&b_{22}
%     \end{vmatrix}
%     +
%     \begin{vmatrix}
% 	b_{11}&a_{12}\\
% 	b_{21}&a_{22}
%     \end{vmatrix}
    .
\end{equation*}
Properties: $*$ is a bilinear product, symmetric and, for all $A,C$,
\begin{eqnarray}
    A*A=2\det A,A*C=A^{\dag}*C^{\dag},\nonumber\\
    \overline{A}C\overline{A}=
    (C*A)\overline{A}-\det(A)\overline{C}.\label{ovABovA=}
\end{eqnarray}

The reference flux-to-source operator has a stable subspace and the
eigenvalue scale is arbitrary.  Therefore, it is no restriction to
take
\begin{equation}
    A=\begin{pmatrix}0&0\\0&-1\end{pmatrix}.
    \label{A=}
\end{equation}
Property: for $A$ defined by (\ref{A=}) and for all $B,C$,
\begin{equation}
    \overline{A}C\overline{B}+\overline{B}C\overline{A}=
    (C*B)\overline{A}+b_{11}\overline{C}-c_{11}\overline{B}.
    \label{ovACovB+ovBCovA=}
\end{equation}

Let $Q=\begin{pmatrix}q_{1}\\q_{2}\end{pmatrix}$ and adjointly
$Q^{\dag}=\begin{pmatrix}q^{\dag}_{1}\\q^{\dag}_{2}\end{pmatrix}$.
\begin{align}
    -\det L_{2}(\epsilon,z')&=-\det(B)z^{\prime 2}-\det(C)\epsilon^{2}-
    B*Cz'\epsilon +b_{11}z'+ c_{11}\epsilon,
    \label{detL2=}\\
    -\det L_{2}\Phi^{*}(\epsilon,z')&=
    (\overline{A}+z'\overline{B}+\epsilon\overline{C})Q,
    \label{detL2P=}\\
    -\det L_{2}R^{*}(\epsilon,z')&=-q^{\dag}_{1}q_{1}+
    \<Q^{\dag}|\overline{B}Q\>z'+\<Q^{\dag}|\overline{C}Q\>\epsilon,
    \nonumber\\
    -\underline{\det L_{2}}(\epsilon)&=-q^{\dag}_{1}q_{1}+
    \<Q^{\dag}|\overline{B}Q\>\underline{z}'+
    \<Q^{\dag}|\overline{C}Q\>\epsilon.
    \label{detL2_=}
\end{align}

I assume, for simplicity, $b_{11}\neq 0$ and
\begin{equation}
    0=\det B=\det C=B*C,
    \label{det=*=0}
\end{equation}
which makes the control variable response linear.  (\ref{det=*=0})
still leaves five degrees of freedom in $B,C$, out of eight generally.

With the new variable \[z\equiv z'-\alpha_{1},\alpha_{1}\equiv
\frac{q^{\dag}_{1}q_{1}}{\<Q^{\dag}|\overline{B}Q\>-b_{11}}\] and
(\ref{det=*=0}), (\ref{detL2=}, \ref{detL2P=}, \ref{detL2_=}) yield
\begin{eqnarray}
    -\det L_{2}(\epsilon,z)&=&b_{11}z+c_{11}\epsilon+b_{11}\alpha_{1}, 
    \nonumber\\
    R^{*}(\epsilon,z)&=&\frac{\<Q^{\dag}|\overline{B}Q\>z+
    \<Q^{\dag}|\overline{C}Q\>\epsilon+
    \alpha_{1}\<Q^{\dag}|\overline{B}Q\>-q^{\dag}_{1}q_{1}}
    {b_{11}z+c_{11}\epsilon+b_{11}\alpha_{1}},
    \label{R*(e,z)=}\\
    \underline{z}(\epsilon)&=&\alpha_{2}\epsilon, 
    \alpha_{2}\equiv-\frac{\<Q^{\dag}|\overline{C}Q\>-c_{11}} 
    {\<Q^{\dag}|\overline{B}Q\>-b_{11}}, 
    \nonumber
\end{eqnarray}
% hypotheses supplementaires pour bijectivite z<->\epsilon
\begin{eqnarray}    
    \underline{\Phi}^{*}(\epsilon)&=&
    \frac{\alpha_{3}}{1-\epsilon\alpha_{4}}
    (\overline{A}+\alpha_{1}\overline{B}+
    \epsilon(\overline{C}+\alpha_{2}\overline{B}))Q,
    \label{P*=}\\
    \alpha_{3}&\equiv&\frac{1}{b_{11}\alpha_{1}},
    \alpha_{4}\equiv\frac{b_{11}\<Q^{\dag}|\overline{C}Q\>-
    c_{11}\<Q^{\dag}|\overline{B}Q\>}{b_{11}q^{\dag}_{1}q_{1}},\nonumber\\
    \alpha_{1}\alpha_{4}+\alpha_{2}&=&-\frac{c_{11}}{b_{11}}.
    \label{1.4+2=}
\end{eqnarray}

To avoid the determination of primitives, I will rather check the
differential form of (\ref{Z1+Z2=01}):
\begin{equation}
    -\<\underline{\Phi}^{\dag*}|C\underline{\Phi}^{*}\>(\epsilon)
    \overset{?}{=} 
    \partial_{2}R^{*}(\epsilon,\underline{z}) 
    \frac{d\underline{z}}{d\epsilon}(\epsilon).
    \label{-w1=d/R/d/e}
\end{equation}

The r.  h. s.  of (\ref{-w1=d/R/d/e}) is evaluated by applying on 
(\ref{R*(e,z)=})
\begin{equation}
    \forall(A,\det A\neq 0),\forall(\underline{z}, 
    \frac{a_{11}\underline{z}+a_{12}}{a_{21}\underline{z}+a_{22}}=1), 
    \frac{d}{dz}\frac{a_{11}z+a_{12}}{a_{21}z+a_{22}}(\underline{z})=
    \frac{(a_{11}-a_{21})^{2}}{\det A}: 
    \nonumber
\end{equation}
% OK
\begin{equation}
    \partial_{2}R^{*}(\epsilon,\underline{z})
    \frac{d\underline{z}}{d\epsilon}(\epsilon)=-
    \frac{(\<Q^{\dag}|\overline{B}Q\>-b_{11})
    (\<Q^{\dag}|\overline{C}Q\>-c_{11})}
    {b_{11}q^{\dag}_{1}q_{1}(1-\epsilon\alpha_{4})},    
    \frac{1}{1-\epsilon\alpha_{4}}=\sum_{n=0}^{\infty}
    \alpha_{4}^{n}\epsilon^{n}.
    \label{w2=}
\end{equation}

The l.  h. s.  of (\ref{-w1=d/R/d/e}) involves the differential weight
of the exciting variable:
\begin{equation}
    \<\underline{\Phi}^{\dag*}|C\underline{\Phi}^{*}\>(\epsilon)=
    w_{1}(\epsilon)=Z_{1}'(\epsilon)=
    \sum_{n=0}^{\infty}\delta T_{n}\epsilon^{n}.
    \nonumber
\end{equation}
The $\delta T_{n}$ are given by (\ref{<dT>n=}), which appears directly
by substituting in
$\<\underline{\Phi}^{\dag*}|C\underline{\Phi}^{*}\>$ the power series
of the perturbed flux (\ref{P*=}) (and adjointly).  The flux and
weight power series coefficients are
\begin{xalignat*}{2}
    \underline{\Phi}_{0}&=
    \alpha_{3}(\overline{A}+\alpha_{1}\overline{B})Q,&
    \underline{\Phi}^{\dag}_{0}&=\alpha_{3}
    (\overline{A}^{\dag}+\alpha_{1}\overline{B}^{\dag})Q^{\dag},
    \nonumber\\
    \underline{\Phi}_{1}&=\alpha_{3}
    (\alpha_{4}\overline{A}-\frac{c_{11}}{b_{11}}\overline{B}+\overline{C})Q,&
    \underline{\Phi}^{\dag}_{1}&=\alpha_{3}
    (\alpha_{4}\overline{A}^{\dag}-
    \frac{c_{11}}{b_{11}}\overline{B}^{\dag}+\overline{C}^{\dag})Q^{\dag},
    \nonumber\\
    \forall p\ge 1,
    \underline{\Phi}_{p}&=\alpha_{4}^{p-1}\underline{\Phi}_{1},&
    \underline{\Phi}^{\dag}_{p}&=\alpha_{4}^{p-1}\underline{\Phi}^{\dag}_{1}.
    \nonumber
\end{xalignat*}
\begin{align}
    \delta T_{0}&=\<\Phi^{\dag}_{0}|C\Phi_{0}\>,
    \nonumber\\
    \forall n\ge 1,\delta T_{n}&=\alpha_{4}^{n-1} 
    (\<\Phi^{\dag}_{0}|C\Phi_{1}\>+\<\Phi^{\dag}_{1}|C\Phi_{0}\>)+
    \alpha_{4}^{n-2}(n-1)\<\Phi^{\dag}_{1}|C\Phi_{1}\>.
    \label{<dT>n=1}
\end{align} 

\begin{displaymath}
    \delta T_{0}=\alpha_{3}^{2}Q^{\dag}(\overline{A}C\overline{A}+
    \alpha_{1}(\overline{A}C\overline{B}+\overline{B}C\overline{A})+
    \alpha_{1}^{2}\overline{B}C\overline{B})Q.
\end{displaymath}
From (\ref{ovABovA=}, \ref{ovACovB+ovBCovA=}, \ref{det=*=0}), and
after a few lines of algebra,
% OK
\begin{equation}
    \delta T_{0}=\frac{(\<Q^{\dag}|\overline{B}Q\>-b_{11})
    (\<Q^{\dag}|\overline{C}Q\>-c_{11})}{q^{\dag}_{1}q_{1}b_{11}}.
    \label{<dT>0=}
\end{equation}
\begin{equation}
    \begin{split}
	\<\Phi^{\dag}_{1}|C\Phi_{1}\>&=\alpha_{3}^{2}\<Q^{\dag}|
	(\alpha_{4}\overline{A}-\frac{c_{11}}{b_{11}}\overline{B}+
	\overline{C})C
	(\alpha_{4}\overline{A}-\frac{c_{11}}{b_{11}}\overline{B}
	+\overline{C})Q\>\\
	&=\<Q^{\dag}|(\alpha_{4}^{2}\overline{A}C\overline{A}+
	(\frac{c_{11}}{b_{11}})^{2}\overline{B}C\overline{B}-
	\alpha_{4}\frac{c_{11}}{b_{11}}
	(\overline{A}C\overline{B}+\overline{B}C\overline{A}))Q\>=0.
    \end{split}
    \label{<P1, CP1>=0}
\end{equation}
Taking (\ref{<P1, CP1>=0}) in (\ref{<dT>n=1}) shows that $w_{1}$ is a
geometric series.  The ratio $\alpha_{4}$ and the first term given by
(\ref{<dT>0=}) are consistent with (\ref{w2=}, \ref{-w1=d/R/d/e}). 
The radius of convergence is $\alpha_{4}^{-1}$.  Other results can be
checked: the spectral properties of $L(z)$, the recursion relations on
perturbation series coefficients\ldots

% The eigenvalues, as functions of the control variable, lie on a
% hyperbola (a parabola if $\operatorname{tr}B=0$).  Is always possible
% to find an interval on which $L(z)$ has two distinct eigenvalues,
% hence the existence of a stable subspace and spectral separation.

%\newpage
\bibliographystyle{unsrt}
\bibliography{biblio}

\begin{thebibliography}{1}

\bibitem{Vulikh}
B.~Z. Vulikh.
\newblock {\em Introduction to functional analysis for scientists and
  technologists}.
\newblock Pergamon Press, 1963.

\bibitem{Bell}
George~I. Bell and Samuel Glasstone.
\newblock {\em Nuclear Reactor Theory}.
\newblock Robert E. Krieger Publishing Co., inc., Malabar, Fl., USA, 1970.

\bibitem{Brillouin}
L\'eon Brillouin.
\newblock {\em Les Tenseurs en M\'ecanique et en \'Elasticit\'e}.
\newblock Masson, Paris, 1938.

\bibitem{Greenspan1}
Ehud Greenspan.
\newblock Developments in perturbation theory.
\newblock {\em Advances in Nuclear Science and Technology}, 9:181--268, 1976.

\bibitem{WeinbergWigner}
Alvin~M. Weinberg and Eugene~P. Wigner.
\newblock {\em The Physical Theory of Neutron Chain Reactors}.
\newblock The University of Chicago Press, 1958.

\bibitem{Bender}
Carl~M. Bender and Steven~A. Orszag.
\newblock {\em Advanced Mathematical Methods for Scientists and Engineers}.
\newblock Mc Graw-Hill Book Company, 1978.

\bibitem{Foell}
W.~K. Foell.
\newblock {\em Small-sample reactivity measurements in nuclear reactors}.
\newblock American Nuclear Society, 1972.

\end{thebibliography}
\end{document}